\documentclass[12pt]{article}
\pdfoutput=1
\usepackage[font=small, labelfont=bf]{caption}
\usepackage{graphicx}
\usepackage{amsmath}
\usepackage{clay}
\usepackage{amssymb}
\usepackage{amsthm}
\usepackage{longtable}
\usepackage{slashed}
\usepackage{verbatim}
\usepackage{cite}
\usepackage{enumerate}
\usepackage{subfig}  
\usepackage{soul}
\usepackage{url}
\usepackage[usenames,dvipsnames]{xcolor}
\usepackage{tikz-feynman}
\tikzfeynmanset{compat=1.0.0}
\usepackage{hyperref}
\hypersetup{colorlinks=true}
\hypersetup{linkcolor=black}
\hypersetup{citecolor=black}
\hypersetup{urlcolor=black}
\usepackage{setspace}
\usepackage{multirow}
\usepackage{wrapfig}
\usepackage{verbatim}
\numberwithin{equation}{section}

\newcommand{\Tr}{{\rm Tr}}

\newcommand{\calC}{\mathcal{C}}
\newcommand{\calM}{\mathcal{M}}
\newcommand{\MpD}{M_{\textrm{Pl};5}}
\newcommand{\Mpd}{M_{\textrm{Pl}}}
\newcommand{\Mpl}{M_{\textrm{Pl}}}

\newcommand{\rbreak}{r_{\text{\st{sym}}}}

\newcommand{\eKK}{e_{\textrm{KK}}}
\newcommand{\LQG}{\Lambda_{\rm QG}}

\theoremstyle{definition}

\newcommand{\beq}{\begin{eqnarray}}
\newcommand{\eeq}{\end{eqnarray}}

\usepackage{titlesec}

\date{February 2022}
\title{Generalized Symmetry Breaking Scales \\ and Weak Gravity Conjectures}

\institution{Chicago}{\centerline{${}^{1}$Enrico Fermi Institute and Kadanoff Center for Theoretical Physics, University of Chicago}}
\institution{Tokyo}{\centerline{${}^{2}$Department of Physics, The University of Tokyo}}
\institution{Berkeley}{\centerline{${}^{3}$Department of Physics, University of California, Berkeley}}

\authors{Clay C\'ordova\worksat{\Chicago}\footnote{e-mail: {\tt clayc@uchicago.edu}},
Kantaro Ohmori\worksat{\Tokyo}\footnote{e-mail: {\tt kant.ohmori@gmail.com}},
and Tom Rudelius\worksat{\Chicago}\footnote{e-mail: {\tt rudelius@berkeley.edu}}}

\begin{document}
\abstract{We explore the notion of approximate global symmetries in quantum field theory and quantum gravity. We show that a variety of conjectures about quantum gravity, including the weak gravity conjecture, the distance conjecture, and the magnetic and axion versions of the weak gravity conjecture can be motivated by the assumption that generalized global symmetries should be strongly broken within the context of low-energy effective field theory, i.e.\ at a characteristic scale less than the Planck scale where quantum gravity effects become important.  For example, the assumption that the electric one-form symmetry of Maxwell theory should be strongly broken below the Planck scale implies the weak gravity conjecture.  Similarly, the violation of generalized non-invertible symmetries is closely tied to analogs of this conjecture for non-abelian gauge theory.  This reasoning enables us to unify these conjectures with the absence of global symmetries in quantum gravity.}
\maketitle

\tableofcontents

\section{Introduction}
\label{sec:intro}

Many recent works have explored the interplay between low-energy effective field theory and quantum gravity in search of necessary conditions for realizing a given effective field theory as the low-energy limit of a UV complete theory of gravity \cite{Vafa:2005ui}.

A number of these necessary conditions have been conjectured. A long-standing folk theorem holds that exact global symmetries are forbidden in a consistent theory of quantum gravity \cite{Banks:2010zn}. The completeness hypothesis holds that there must exist states in every representation of the gauge group \cite{polchinski:2003bq}. The weak gravity conjecture holds that gravity must be the weakest force, i.e., there must exist a particle whose gauge charge is larger than its mass in Planck units \cite{Arkanihamed:2006dz}. The higher-form weak gravity conjecture extends this idea from point particles charged under ordinary gauge fields to general branes charged under higher-form gauge fields, and the magnetic weak gravity conjecture extends it from electrically-charged objects to magnetically-charged ones.  The distance conjecture holds that an infinite tower of states must become exponentially light at large distances in moduli space \cite{Ooguri:2006in}. Reviews of these conjectures may be found in \cite{Brennan:2017rbf, Palti:2019pca, vanBeest:2021lhn, Grana:2021zvf, Harlow:2022gzl}.

The evidence for these conjectures varies. The absence of global symmetries is on the firmest footing, and arguments in its favor have been given on the basis of black hole physics \cite{Banks:2010zn}, string theory \cite{banks:1988yz}, and holography \cite{Harlow:2018tng}. More recently, it has been demonstrated that including Euclidean wormholes in the gravitational path integral leads to explicit symmetry global symmetry violation in two-dimensional theories of gravity \cite{Harlow:2020bee, Chen:2020ojn, Yonekura:2020ino, Hsin:2020mfa} complementing previous work anticipating these results \cite{sannan:1986tz, abbott:1989jw}. 

There is also strong evidence against higher-form global symmetries in quantum gravity. $p$-form symmetries are symmetries whose charged operators live on manifolds of dimension $p$ \cite{Gaiotto:2014kfa}. By compactifying a theory in $d$ dimensions with a $p$-form global symmetry on a $p$-torus $T^p$, one obtains a $(d-p)$-dimensional theory with an ordinary (0-form) global symmetry \cite{Banks:2010zn}. The authors of \cite{Harlow:2018tng} used this fact to argue against $p$-form global symmetries in AdS$_d$ with $d \geq p+3$.
  
The evidence for the weak gravity conjectures is also formidable. For instance, they are supported by a large class of examples in string theory and Kaluza-Klein theory \cite{Arkanihamed:2006dz, Heidenreich:2016aqi, Montero:2016tif, Palti:2017elp, Lee:2018spm, Lee:2018urn}. Additional arguments for the weak gravity conjecture have also been given on the basis of black hole entropy \cite{Cottrell:2016bty, Fisher:2017dbc, Cheung:2018cwt}, scattering amplitudes \cite{cheung:2014ega, Andriolo:2018lvp, Charles:2019qqt, Arkani-Hamed:2021ajd}, cosmic censorship \cite{Crisford:2017gsb, Horowitz:2019eum}, and holography \cite{Harlow:2015lma, Harlow:2018tng, Montero:2018fns}. Similarly, the axion weak gravity conjecture has been verified by examples in \cite{banks:2003sx, rudelius:2014wla, Delafuente:2014aca, rudelius:2015xta, Montero:2015ofa, Brown:2015iha, Heidenreich:2015nta, Heidenreich:2015wga, Brown:2015lia, long:2016jvd, Conlon:2016aea}, while the distance conjecture has been discussed in \cite{Baume:2016psm, Klaewer:2016kiy, Valenzuela:2016yny, Grimm:2018ohb}, among other works.  For further discussion see the recent review \cite{Harlow:2022gzl}.

Despite the evidence in support of these conjectures, their underlying motivation remains mysterious. The weak gravity conjecture was originally motivated by a desire to forbid stable extremal black holes, but it is unclear why such stable black holes should be forbidden in the first place. Other conjectures, like the distance conjecture, were formulated primarily based on examples in string theory. A more fundamental understanding of these conjectures is desirable.

Furthermore, the relationships between these various conjectures are nebulous. Recent work has suggested a number of connections between them. For instance, the emergence proposal of \cite{Harlow:2015lma, Heidenreich:2017sim, Grimm:2018ohb, Heidenreich:2018kpg} offers a possible unifying principle between the weak gravity conjecture and the distance conjecture: beginning with a strongly coupled gauge field in the UV, a weakly coupled gauge theory emerges in the IR by integrating out a tower of charged particles that satisfy the weak gravity conjecture bound \cite{Heidenreich:2017sim}. Analogously, beginning with a strongly coupled scalar field in the UV, a weakly coupled scalar field theory emerges in the IR by integrating out a tower of light particles, as stipulated by the distance conjecture \cite{Grimm:2018ohb, Heidenreich:2018kpg}.

Connections between the absence of higher-form global symmetries and the completeness hypothesis have been developed at length in \cite{Banks:2010zn, Harlow:2015lma, Harlow:2018tng, Rudelius:2020orz, Heidenreich:2021tna, Casini:2021zgr}. In a pure gauge theory, Wilson lines are charged under a 1-form symmetry.\footnote{In the case of a finite non-abelian group or a disconnected gauge group, this 1-form symmetry may be a ``non-invertible'' symmetry. See \cite{Hsin:2021qiy} and references therein for more on this topic.} In the presence of dynamical charged matter, Wilson lines can open into pairs of charged particles and hence the associated 1-form symmetry is broken.  With sufficiently many matter fields in different representations of the gauge group, the one-form symmetry will be broken entirely. In this way, the absence of higher-form symmetries in quantum gravity is intimately related to the completeness hypothesis. The completeness hypothesis also appears necessary for factorization of the Hilbert space of the thermofield double, as factorization requires that a Wilson line stretching from one boundary to the other must be able to split open at a pair of charged particles \cite{Harlow:2015lma}.

In this work, we will see that many of these conjectures can be succinctly unified by quantifying the strength of symmetry violating effects.  We will study general symmetries which appear in low-energy effective field theories.  These symmetries are typically approximate and are broken by high energy effects.  Insisting that approximate global symmetries--including approximate generalized global symmetries--cannot exist at the Planck scale, is nearly equivalent to many of the most well-supported conjectures about quantum gravity.  Note that this is stronger than assuming that symmetries are violated by small quantum gravity effects.  Instead, having all approximate symmetries badly violated below the Planck scale means that the symmetry breaking processes must become important within the effective field theory itself.

A central example is pure abelian gauge theory which features a 1-form symmetry where the associated charged objects are Wilson lines. To badly break this symmetry below the Planck scale, it is not sufficient merely to satisfy the completeness hypothesis with particles of every charge. If the charged states are much heavier than the Planck scale, their symmetry-violating effects will be negligible at that scale. Thus, the charged states in question must be sufficiently light. Furthermore, to effectively screen the charge of the Wilson line, they must carry a sufficiently large charge, or else there must be a sufficiently large number of these light charged species. Morally speaking, the requirement of light particles of large charge is simply the statement of the weak gravity conjecture. More precisely, as we will see below, the 1-form global symmetry is badly broken precisely when the gauge theory becomes strongly coupled, and the requirement that this should occur within effective field theory due to a tower of light charged states is essentially the content of the emergence proposal. This in turn requires a tower of charged particles satisfying the weak gravity conjecture bound \cite{Heidenreich:2017sim}.

A similar argument can be made for the distance conjecture: a free, massless scalar field theory features a 0-form symmetry under which the scalar field shifts by a constant. If the scalar is coupled to a fermion with a Yukawa coupling, this symmetry will be broken. However, it may remain as an approximate symmetry unless the scalar field becomes strongly coupled. This will occur below the Planck scale provided the scalar couples to a tower of particles of increasing mass, as required by the distance conjecture, and integrating out this tower of particles leaves a weakly-coupled scalar field theory in the IR, as expected from the emergence proposal \cite{Grimm:2018ohb, Heidenreich:2018kpg}.

Similar statements can be made for other versions of the weak gravity conjecture as well. In particular, we will see that in certain situations, the magnetic weak gravity conjecture and the axion weak gravity conjecture can also be motived by the statement that (higher-form) symmetries should be badly broken below the Planck scale. We will also argue that a non-abelian analog of the weak gravity conjecture follows from assumptions about violation of certain ``non-invertible'' higher-form symmetry defects.

Our work relies on several assumptions whose justification is a target for future work. Most notably, we do not claim to have a compelling argument for why approximate symmetries must be badly broken within the low-energy effective field theory. However, we consider the connections developed here between this principle and various quantum gravity conjectures to be a tantalizing clue that this principle is on the right track and deserving of future study.

The remainder of this paper is organized as follows. In section \ref{sec:Global}, we review the notion of a higher-form global symmetry and arguments against global symmetries in quantum gravity. We introduce the notion of an approximate global symmetry and quantify the symmetry breaking scale. In section \ref{sec:Swampland}, we show how the requirement that 1-form symmetries should be badly broken below the Planck scale can be used to motivate the weak gravity conjecture and the magnetic weak gravity conjecture, and we comment on a non-abelian extension of our arguments. In section \ref{sec:SHIFT}, we likewise show that badly breaking a 0-form symmetry below the Planck scale leads to the distance conjecture and the axion weak gravity conjecture. In section \ref{sec:CONC}, we end with conclusions and directions for future research.

\section{Approximate Global Symmetries}\label{sec:Global}

\subsection{Approximate Global Symmetries and Black Holes}\label{sec:approx}

A widely-believed folk theorem is that exact continuous global symmetries are inconsistent with quantum theories of gravity \cite{Misner:1957mt, Banks:2010zn}. This idea has recently found strong support in holography \cite{Harlow:2018tng} and has even been derived in certain low-dimensional toy models from the physics of wormholes \cite{sannan:1986tz, abbott:1989jw, Harlow:2020bee, Chen:2020ojn, Yonekura:2020ino, Hsin:2020mfa}.

Let us revisit the original argument motivating this conclusion, focusing on the simplest example of a $U(1)$ global symmetry (see e.g.\ \cite{Banks:2010zn}).  Consider a theory with a field $\phi$ with charge $q$. A  large number of $\phi$ particles with sufficient density can collapse to form a black hole with charge $Nq$ for $N>>1$.  Quantum mechanically, this black hole will decay via the Hawking evaporation process.  In so far as this process may be treated semiclassically, i.e.\ for large enough black holes, the radiation appears thermal and hence does not on average modify the charge of the black hole.  This continues until the black hole becomes too small and Hawking's computation becomes unreliable.  

As a result, we are left with a small black hole with Schwarzschild radius $r_S$ of order the Planck length $l_{\rm Pl}$ with a large global symmetry charge.  Moreover, for sufficiently large $N$ above, this black hole will be kinematically unable to shed its large charge by emitting charged particles and therefore gives rise to a stable Planck-sized remnant.  As this analysis holds for all black holes of sufficiently large charge, we further deduce that there are an infinite number of distinct Planck sized remnants, labeled by their total charge.  However, since this is a a global symmetry charge, these various charged remnants do not source any long-range fields and are hence indistinguishable to distant observers.  In other words, the black holes with different global charges all correspond to a single black hole macrostate, violating the Bekenstein-Hawking formula for black hole entropy.

Consider now the same argument but applied not to an exact global symmetry, but an approximate global symmetry.  Here, by an approximate global symmetry, we mean a conservation law that is valid at long distances and is violated at some ultraviolet scale $\Lambda$.  Note that according to the general reasoning above, all symmetries of low-energy effective field theories in models with dynamical gravity should be interpreted as approximate.  For instance, in the standard model $U(1)_{B-L}$ is a symmetry at the level of the renormalizable low-energy Lagrangian, but it is natural to expect it is violated (at least) by high energy gravitational processes.  

In general, in this approximate global symmetry scenario, we no longer expect exactly stable black hole remnants, since the end products of the Hawking process may be able to decay via symmetry-violating channels.  However, if the symmetry breaking scale $\Lambda$ is parametrically larger than the Planck scale, charge violating decays are suppressed and we will again be left a large collection of long-lived black hole states, in tension with entropy bounds.  

The above reasoning motivates our general conjecture that in quantum theories of gravity, the scale $\Lambda$ of symmetry violation must be at or below the Planck scale: 
\begin{equation}\label{symconj}
\Lambda \lsim M_{\text{Pl}}~.
\end{equation}
In this case, processes that violate the approximate symmetry will become important before the black hole becomes Planckian in size, $r_S \sim l_{\rm Pl}$, and there will be no meaningful way to assign an approximate charge to the remnant.\footnote{In models with a large number $N$ of light species, the right-hand side of equation \eqref{symconj} should be replaced by the scale \begin{equation}
\Lambda_{\rm QG} \sim N^{-\frac{1}{d-2}} M_{\textrm{Pl};d}~,
\label{eq:LQG}
\end{equation} where gravity becomes strongly coupled ($d$ is the spacetime dimension)  \cite{Dvali:2007wp}. At this scale, loop contributions to the graviton propagator rival the tree-level value, and the semiclassical description of gravity breaks down. } This assumption is closely connected to the idea that the underlying symmetry violating processes are non-gravitational and occur within the context of a low-energy effective field theory.  

By considering black branes charged under higher-form gauge fields, or by first compactifying a theory with an approximate $p$-form global symmetry on a $p$-torus to obtain an approximate 0-form symmetry, we expect that higher-form symmetries should also be badly broken at or below the Planck scale. Below we will show that this idea is closely related to the weak gravity conjecture and to a variety of other proposed consistency conditions in quantum gravity.

\subsection{Global Symmetry Breaking}\label{sec:gsb}

Let us better quantify the notion of the symmetry breaking scale.  As described in section \ref{sec:intro}, a key role will be played by higher-form global symmetries and their breaking.  Our treatment follows the general discussion of these symmetries in \cite{Gaiotto:2014kfa}.

A continuous, abelian, $p$-form global symmetry in $d$ dimensions is associated to a closed $d-p-1$-form current operator $J$.\footnote{The Hodge dual $*J$ is then a $p+1$-form operator, which is conserved.} By integrating this operator over closed $(d-p-1)$-manifolds $\calM^{(d-p-1)}$ we create symmetry operators 
\begin{equation}
U_{\alpha}(\calM^{(d-p-1)})\equiv \exp{\left(i\alpha\int_{\calM^{(d-p-1)}} J\right)}~.
\end{equation}
The most important feature of these symmetry operators is that they they are topological: since $J$ is closed, small changes in the manifold $\calM^{(d-p-1)}$ do not modify the correlation functions of the operator $U_{\alpha}(\calM^{(d-p-1)})$.  Additionally, the group law is encoded in the fusion algebra of these operators for different phases $\alpha$.

\begin{figure}[t]
	\centering
	\begin{tikzpicture}[scale = .6,thick,baseline = 0]
		\draw (0,0) coordinate (0) arc (-180:180:2 and 1);
		\node (U) at ($(0)+(-.5,.5)$) {$U_g$};
		\coordinate (c) at ($(0)+(2,0)$);
		\draw[line width=5pt,white] (0) ++ (1,-.7) .. controls ($(c)+(-.5,0)$) and ($(c)+(1,1)$) .. ++ (3,4);
		\draw (0) ++ (1,-.7) coordinate (1) .. controls ($(c)+(-.5,0)$) and ($(c)+(1,1)$) .. ++ (3,4) coordinate(2);
		\draw (1) ++ (-.2,-.25) -- ++(-.8,-1);
		\node (L) at ($(2)+(-1.4,-.2)$) {$V(\calC^{(p)})$};
	\end{tikzpicture}
	$=  e^{i \alpha q} \times $
	\begin{tikzpicture}[scale = .6,thick, baseline = 0]
		\draw (0,0) coordinate (0);
		\coordinate (c) at ($(0)+(2,0)$);
		\draw (0) ++ (1,-.7) coordinate (1) .. controls ($(c)+(-.5,0)$) and ($(c)+(1,1)$) .. ++ (3,4) coordinate(2);
		\draw (1) -- ++(-.2,-.25) -- ++(-.8,-1);
		\node (L) at ($(2)+(-1.4,-.2)$) {$V(\calC^{(p)})$};
	\end{tikzpicture}
	\caption{The action of a $p$-form symmetry operator $U_\alpha$  on a $p$-dimensional extended operator $V( \calC^{(p)} )$. When $U_\alpha$ surrounds $\calC^{(p)}$, it acts by a phase $\exp ({i \alpha q })$. }
	\label{fig:p-form_action}
\end{figure}
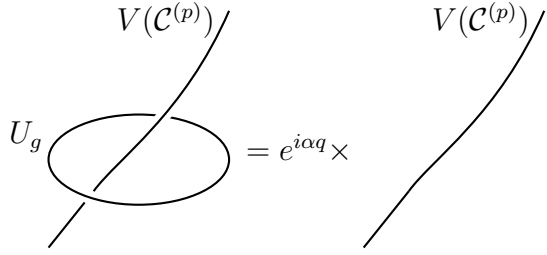

Although small deformations do not modify $U_{\alpha}(\calM^{(d-p-1)})$, the correlation functions do change when $\calM^{(d-p-1)}$ is deformed across an operator $V(\calC^{(p)})$ that is charged under the symmetry.  Such charged operators are supported on $p$-dimensional manifolds $\calC^{(p)}$. Taking $\calM^{(d-p-1)} = S^{d-p-1}$ to be a sphere surrounding $\calC^{(p)}$, we have
\begin{equation}
U_{\alpha}(S^{d-p-1}) V(\calC^{(p)}) = \exp{\left(i\alpha q\right)} V(\calC^{(p)})~, 
\label{eq:flux}
\end{equation}
where $q$ is the charge of the operator $V$ (see Figure \ref{fig:p-form_action}.)  

Consider now the situation where the symmetry is not exact but is broken by small effects.  In this case, the operators $U_{\alpha}(\calM^{(d-p-1)})$ will still exist, but they will no longer be topological, since the associated current $J$ is no longer conserved.  This means that in \eqref{eq:flux}, the phase on the right-hand side will become dependent on the radius $r$ of the sphere $S^{d-p-1}$. This can be encoded by allowing the effective charge $q$ depend on radius: \begin{equation}\label{qdef}
\lim_{\alpha \rightarrow 0}\frac{1}{i\alpha}\log\left(\frac{\langle V^{\dagger}(\calC^{(p)})(\infty)U_{\alpha}(S_r^{d-p-1}) V(\calC^{(p)})(0)  \rangle }{ \langle V^{\dagger}(\calC^{(p)})(\infty)V(\calC^{(p)})(0) \rangle}\right)\equiv q(r)~.
\end{equation}
Here the charged operator is inserted at a conveniently chosen interior region of spacetime, denoted schematically as $0,$ and the conjugate operator $V^{\dagger}$ is inserted at a distant location, schematically denoted as $\infty$.  We choose these configurations such that the two point function in the denominator above is reflection positive and hence cannot vanish.

If the symmetry becomes conserved in the infrared, then for a sphere of large radius the effective charge tends toward the IR value of the charge: $q(r)\rightarrow q$.  In particular, since the symmetry operators become topological in the IR, the large radius value of $q(r)$ is insensitive to small changes in the radius $r$.

To quantify the notion of symmetry breaking in the ultraviolet, we examine the radial dependence of the effective charge $q(r)$ at short distances.  Specifically, we will say that the symmetry is badly broken at a distance scale $r_{\text{\st{sym}}}$ (related to an energy scale by $r_{\text{\st{sym}}}\sim \Lambda^{-1}$), if the effective charge varies with distance as 
\begin{equation}
\frac{1}{q_\infty}\left(\frac{ d }{  d \log r}\right)q(r)\Big|_{r=r_{\text{\st{sym}}}} \sim O(1)~.
\label{eq:GSB}
\end{equation}
Where in the above, the normalization by $q_{\infty}$, the IR value of the charge, makes the scale $r_{\text{\st{sym}}}$ insensitive to the IR charge of the particular operator selected to probe symmetry breaking effects.
 
Let us elaborate on possible power law ambiguities in the effective charge $q(r)$ and the UV sensitivity of the definition \eqref{eq:GSB}.  To probe the short distance behavior of $q(r)$ requires extrapolating the current operator $J$ to the ultraviolet. At long distances, the  operator $J$ is distinguished by being conserved, but at short distances this conservation breaks down, and hence there may be no canonical definition of the current.  Consider for concreteness a current for an ordinary symmetry (i.e., a closed $(d-1)$-form).  We can modify the current $J$ at short distances as:
\begin{equation}\label{jmod}
\delta J = \frac{\mathcal{O}}{\Lambda^{\Delta_{\mathcal{O}}-d+1}}~,
\end{equation}
where $\mathcal{O}$ is any local operator transforming as a $d-1$ form, $\Delta_{\mathcal{O}}$ is its scaling dimension, and $\Lambda$ is an ultraviolet cutoff. For $\Delta_{O}$ sufficiently large, this modification drops out at low energies but can modify the properties of $J$ at short distances.  In particular, applying \eqref{qdef} we expect that the above leads to a short distance modification of the effective charge as:
\begin{equation}
\delta q(r)\sim \frac{1}{(r\Lambda)^{\Delta_{\mathcal{O}}-d+1}}~.
\end{equation}
Since these contributions can modify the short distance definition of the charge $q(r)$, we see that the precise scale of symmetry breaking depends in detail on how the effective field theory current operator is extrapolated to the ultraviolet.

In practice, although such ambiguity exists, we expect that the symmetry violating scale cannot be parametrically modified unless the current is modified by an operator with a large coefficient.  Thus our analysis will be robust under generic small deformations.  Our conclusions become more universal, however, when the symmetry breaking effects come from terms in the effective charge $q(r)$ which are non-analytic in $r$ at short distances.  Such non-analyticity is a universal prediction of the effective field theory. We will see this behavior in our analysis of the weak gravity conjecture below.\footnote{Although our primary focus below is on 4d theories similar remarks apply to other dimensions.  In particular when studying higher-form symmetry breaking in general spacetime dimension we will need to contend with the fact that the gauge coupling is classically dimensionful.  Consistent with the discussion above one should define the symmetry breaking scale in general to be where non-analytic corrections to the charge become large. }

\subsubsection{Example: Complex Scalar Field}

It is instructive to illustrate how the formal definition \eqref{eq:GSB} coincides with usual notions of symmetry violation in effective field theory.  
As an example,  let $\varphi$ be a weakly coupled massless complex scalar.  There is a global symmetry that rotates the phase of the field $\varphi$. The current operator is
\begin{equation}
(*J)_\mu = \frac{i}{2} (\varphi^\dagger \partial_\mu \varphi - \varphi \partial_\mu \varphi^\dagger )~.
\end{equation}
The local operator $\varphi^{q}$ carries charge $q$ and hence
\begin{equation}
e^{i \alpha \int_{S^3_r} J} \varphi^{q}(x) = e^{i q \alpha} \varphi^{q}(x)~,
\end{equation}
where the integral is over an $S^3$ of radius $r$ centered at the insertion of $\varphi^{q}$. Passing to expectation values and expanding to linear order in the phase $\alpha$ this gives
\begin{equation}
i\langle  \int_{y\in S^3_r} \epsilon^{\mu\nu\sigma\rho}  d \Sigma_{\mu\nu\sigma} \cdot( \varphi \partial_\rho \varphi^\dagger-\varphi^\dagger \partial_\rho \varphi  )(y) \varphi^{q}(x) \cdots \rangle  = q \langle \varphi^{q}(x) \cdots \rangle ~,
\label{eq:phisymmetry}
\end{equation}
where the unwritten terms $\cdots$ stand for other possible insertions in the correlator.

As in the general discussion above, the fact that the global symmetry is exact implies that this integral is topological: it does not depend on the radius $r$ of the three-sphere.  However, let us now suppose that we add a symmetry-breaking higher-dimension operator to the Lagrangian of the form:
\begin{equation}
\Delta \mathcal{L} = \frac{1}{\Lambda^{n-4}} (\varphi^n + (\varphi^\dagger)^n)~.
\end{equation}
Such an operator breaks the global symmetry, as the current $J$ is no longer conserved:
\begin{equation}
d J \sim  \frac{i }{\Lambda^{n-4}} ((\varphi^\dagger)^{n-1} \varphi -\varphi^\dagger \varphi^{n-1} ) \neq 0~.
\end{equation}
At energies far below $\Lambda$, this operator is negligible, and there is an approximate global symmetry. Intuitively, we expect that in the presence of such a potential, the phase rotation symmetry of $\varphi$ is badly broken at the scale $\Lambda$.  We would like to verify that using the more quantitative definition in \eqref{eq:GSB}.

To do so, we need to compute the corrections to \eqref{eq:phisymmetry} in the presence of the higher-dimension operator. This is easy to understand in terms of Feynman diagrams. For instance, in the absence of the symmetry violating interaction the left-hand side of (\ref{eq:phisymmetry}) can be represented diagrammatically in the case $q=1$ as
\begin{equation}
\includegraphics[width=60mm]{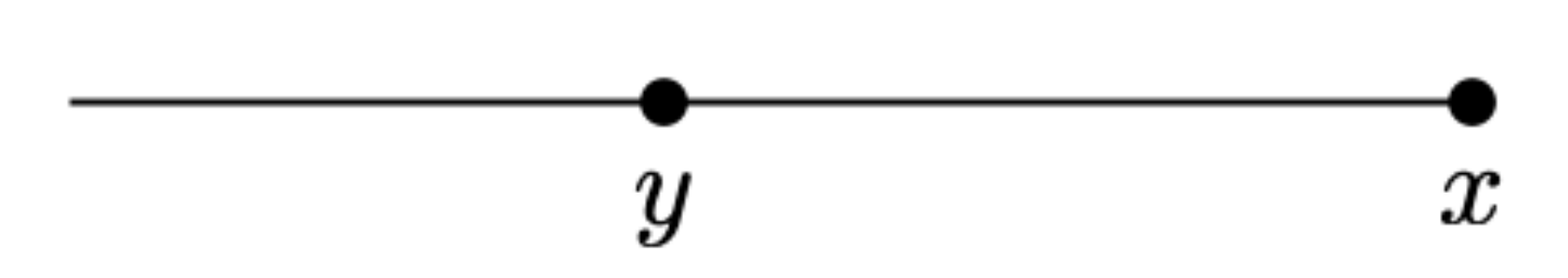}~,
\end{equation}
where $y \in S_r^3$, and after evaluating the diagram we integrate over $y$. Once the symmetry violating operators are added to the Lagrangian the above is corrected by diagrams involving insertions of the interactions.  The leading order correction is a so-called sunrise diagram computed for instance in \cite{Aste:2006iv}:
\begin{equation}
\includegraphics[width=70mm]{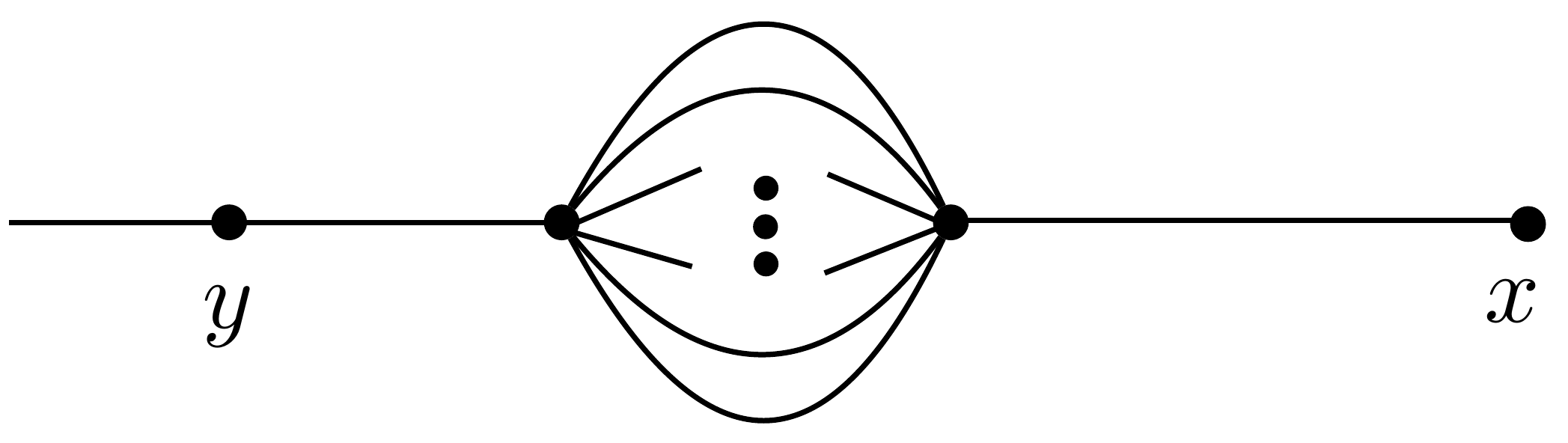}~.
\end{equation}
For free $x$ and $y$, this can be interpreted as a correction to the scalar propagator.  

Working in momentum space and ignoring order one coefficients, the resulting self-energy function $\Pi(p^{2})$ scales as:
\begin{equation}
\Pi(p^2) \sim \left(\frac{p}{\Lambda}\right)^{2n-8}~.
\end{equation}
To compute the variation of the charge $q(r)$ in \eqref{eq:GSB}, we Wick rotate and Fourier transform into position space to obtain the two-point function $G(r)$ and then integrate over an $S^3$ of radius $r$. We have
\begin{equation}
G(r)\sim \frac{1}{r} \int_{-\infty}^\infty dp K_1(pr)\Pi(p^2) \sim \frac{1}{r^2} \frac{1}{(\Lambda r)^{2n-8}} ~,
\end{equation}
where $K_{1}(x)$ is a Bessel function.  The definition \eqref{qdef} then gives:
\begin{equation}
q(r)\sim \int_{S^{3}_{r}} G(r) \sim \frac{1}{(\Lambda r)^{2n-8}}~.
\end{equation}
The general condition \eqref{eq:GSB} gives the symmetry breaking scale $\frac{1}{r_{\text{\st{sym}}}} \sim \Lambda$ as expected on dimensional grounds.

In an effective field theory that arises from a low-energy limit of quantum gravity, one might expect that the absence of global symmetries leads to symmetry violating operators that are suppressed by the quantum gravity scale $\LQG$. From the computation we have just performed, we therefore see that $r_{\text{\st{sym}}}^{-1} \sim \LQG$ for such a theory. In what follows, we will explore analogs of this idea for higher-form global symmetries.

\section{Symmetry Breaking and Weak Gravity Conjectures}\label{sec:Swampland}

In this section, we argue that a number of proposed quantum gravity consistency conditions follow from the statement that approximate global symmetries must be broken by an $O(1)$ amount at or below the quantum gravity scale $\LQG$.  

\subsection{The Weak Gravity Conjecture}\label{sec:TWGC}

The ordinary weak gravity conjecture holds that in an abelian gauge theory in $d$ dimensions with coupling constant $g$, there must exist a charged particle of mass $m$, quantized charge $n \in \mathbb{Z}$, such that
\begin{equation}\label{WGCstate}
\frac{g n}{m} \geq \frac{\gamma_d}{M_{\textrm{Pl};d}^{(d-2)/2}}~,
\end{equation}
where $\gamma_d$ is an $O(1)$ number that depends on $d$, the precise value of which can be found in \cite{Heidenreich:2015nta}. A particle that satisfies the bound is often called ``superextremal.'' Several different lines of evidence suggest that a consistent theory of quantum gravity should have not just one superextremal particle, but rather an infinite tower of increasing mass and charge \cite{Heidenreich:2015nta, Heidenreich:2016aqi, Heidenreich:2017sim, Grimm:2018ohb, Lee:2018spm, Lee:2018urn, Andriolo:2018lvp, Lee:2019xtm,Lee:2019tst, Heidenreich:2019zkl}. This stronger hypothesis is sometimes called the tower weak gravity conjecture or the sublattice weak gravity conjecture, if one further demands that these particles lie on a sublattice of the full charge lattice.

Historically, one way that the weak gravity conjecture has been motivated is by examining the weak coupling limit of the gauge theory, $g\rightarrow 0$.  In that case the gauge fields decouple and intuitively the matter sector develops a global symmetry which is sourced by the gauge fields.  Since exact global symmetries are forbidden in a consistent theory of quantum gravity, one expects that this weak-coupling limit is also forbidden. The inequality \eqref{WGCstate} quantifies this intuition.  Instead, we will motivate the weak gravity conjecture by examining the one-form global symmetries and their pattern of symmetry breaking.

First, let us recall that a free abelian gauge field (neglecting any couplings to matter) has two higher-form global symmetries $U(1)^{(1)}_{e}$ and $U(1)^{(d-3)}_{m}$.  Here, the subscripts $e$ stands for ``electric" while $m$ stands for ``magnetic" and indicate the objects charged under these symmetries.  For $U(1)^{(1)}_{e}$ these are electrically charged Wilson lines, while for $U(1)^{(d-3)}_{m}$ these are magentically charged 't Hooft operators.  The presence of these symmetries can also be seen via the associated currents:\footnote{We work in Euclidean signature with canonically normalized kinetic terms for the gauge fields.  With these conventions the charges of the currents below are integers.}
\begin{equation}\label{emcurrs}
J_{e}=\left(\frac{i}{g}\right)*F, \hspace{.3in}J_{m}=\left(\frac{g}{2\pi}\right)F~,
\end{equation}
which are closed respectively, due to the equation of motion and the Bianchi identity.  

We can formulate these symmetries using the language described in section \ref{sec:gsb}.  For, $U(1)^{(1)}_{e}$, the charged operators are Wilson lines:
\begin{equation}\label{wdef}
W_\gamma(q) =  \exp \left( i g q \oint_{\gamma} A_\mu d x^\mu \right)~, \hspace{.5in}q\in \mathbb{Z}~,
\end{equation}
where the integer charge $q$ is the electric charge of the line.  The symmetry generator associated with a $d-2$-sphere of radius $r$ is given by
\begin{equation}\label{1formabop}
U_{\alpha}(S_r^{d-2}) =  \exp \left(- \frac{\alpha }{g} \int_{S^{d-2}} *F \right)~.
\end{equation}
Plugging this into \eqref{eq:flux}, we see that the symmetry generator acts on a Wilson line at the center of its associated $S^{d-2}$ by
\begin{equation} \label{uweq}
U_{\alpha}(S_r^{d-2}) W_\gamma(q)= \exp(i\alpha q) ~W_\gamma(q)~.
\end{equation}
Analogous expressions hold for the magnetic symmetry acting on 't Hooft operators.

By the general logic described in section \ref{sec:approx}, it is natural to expect that the one-form global symmetries in Maxwell theory must be broken below the Planck scale.  In this section we will focus on the electric one-form symmetry $U(1)_{e}^{(1)}$ and see that its breaking is closely tied to the weak gravity conjecture.  In section \ref{ssec:magnetic}  we will see that breaking the $U(1)_{m}^{(d-3)}$ is closely tied to the magnetic weak gravity conjecture.

As a first analysis of the symmetry $U(1)_{e}^{(1)}$ one might ask if it can simply be violated, as in the case of ordinary symmetries, by deforming free Maxwell theory by symmetry violating operators at short distances.  Surprisingly, we will now argue that this is not the case.  Specifically, as long as no new degrees of freedom are added, the symmetry $U(1)_{e}^{(1)}$ persists under local operator deformations of the effective action.  

To see this more quantitatively consider adding to the Lagrangian a gauge invariant higher-dimension operator:
\begin{equation}\label{opdef}
\delta \mathcal{L}=\frac{1}{\Lambda^{k}} \mathcal{O}(F)~.
\end{equation}
Here our notation indicates that, since $\mathcal{O}(F)$ is gauge invariant, it is a polynomial function in the field strength and spacetime derivatives contracted to form a Lorentz scalar.  For instance, $\mathcal{O}(F)$ could be the famous $|F|^{4}$ operator arising in the Euler-Heisenberg Lagrangian.  By varying the effective action one can compute the equation of motion to find:
\begin{equation}\label{modecur}
d\left(*F +*\frac{2}{\Lambda^{k}}\frac{\partial \mathcal{O}}{\partial F}\right)=0~.
\end{equation}
Notice that the $\frac{\partial \mathcal{O}}{\partial F}$ is a well-defined (gauge invariant) two-form operator, so we can view the above as a conservation equation for a modified electric one-form symmetry current.  Thus as asserted above, local operator modifications of the Maxwell EFT are insufficient to generate the anticipated symmetry violation of $U(1)_{e}^{(1)}$.

Let us instead now turn to coupling the Maxwell theory to massive charge matter fields.  In this case we will indeed find that symmetry violation is possible due to the physics of charge screening.  In the presence of charged matter the equation of motion for the gauge field is modified to: 
\begin{equation}
(d  \, {* F})^{\mu\nu\rho}  \sim   g n \varepsilon^{\mu\nu\rho \sigma} \bar \psi \gamma_{\sigma} \psi~.
\label{eq:EOM}
\end{equation}
Unlike the the analysis of the operator deformation \eqref{opdef}, the $U(1)_{e}^{(1)}$ symmetry current is now genuinely broken.  Far below the mass scale $m$ of the charged matter the fermions freeze out and the electric one-form symmetry is restored.  Our goal is therefore to investigate this approximate one-form symmetry breaking induced by charged matter.  Throughout we work in the effective field theory where the gauge coupling $g$ is small.  In this case, the violation of $U(1)_{e}^{(1)}$ stemming from a single charge particle is suppressed and we can analyze symmetry breaking effects in perturbation theory.

We now proceed with the analysis of the effective charge $q(r)$ outlined in section \ref{sec:gsb}.  Expanding \eqref{uweq} to order $\alpha$ gives
\begin{equation}
\left( \frac{i}{g} \int_{S^{d-2}} *F  \right) W_{\gamma} (q_{\infty})\cdots = q(r) W_{\gamma}(q_{\infty}) \cdots~.
\label{eq:Wardexp}
\end{equation}
Here the terms in $\cdots$ represent other potential insertions in a correlation function. The integer $q_{\infty}$ appearing in the argument of $W_{\gamma} (q_{\infty})$ is the electric charge used to define the line in \eqref{wdef}.  This is also the $U(1)_{e}^{(1)}$ one-form symmetry charge in the long distance limit far below the mass scale of charged particles.  Meanwhile, the desired effective charge, $q(r)$ is simply the integral of the current $*F$ over an $S^{d-2}$ of radius $r$ in the presence of a background Wilson line. 

We can translate this into a more familiar language by taking the Wilson line to be along time at the origin of space.  In that case $*F$ is proportional to the electric field.  The Wilson line sources an effective potential, $V(r)$ and our task is to study the modification of Gauss' law.  
\begin{equation}\label{qrwline}
q(r)=\frac{1}{g^{2}}\int _{S^{d-2}_{r}} \frac{dV(r)}{dr}~,
\end{equation}
This position dependence of $q(r)$ can be understood in terms of the screening of the electric charge due to virtual particles. 

In 4d, these ideas are famously quantified by the Uehling potential \cite{Uehling:1935uj}, which is the effective potential of a point source in the presence of a charged particle of mass $m$, charge $n$, at distances $r \gg 1/m$.  At leading order in the coupling one finds:
\begin{equation}
V(r) = \frac{-g^2q_{\infty}}{4 \pi r} \left( 1 + \frac{n^2 g^2}{16 \pi^{3/2}} \frac{e^{-2 m r}}{ (m r)^{3/2}} + ... \right)~.
\label{eq:largem}
\end{equation}
The first term represents the effective Coulomb potential felt by an observer at infinity. This is smaller than the bare charge due to screening by the particle. At these large distances, however, the effects of the particle are exponentially suppressed, and as a result the charge is approximately conserved.  In particular, evaluating \eqref{qrwline} gives:
\begin{equation}\label{largerq}
q(r)=q_{\infty}\left(1+g^{2}n^{2}\frac{e^{-2m r}}{\sqrt{64\pi^{3}mr}}+\cdots \right)~, \hspace{.2in} mr \gg1~.
\end{equation}

Notice that the position dependence of the effective charge $q(r)$ is analytic for large $r$, its regime of validity.  As in the discussion around \eqref{jmod}, this means that the symmetry breaking effects encoded by $V(r)$ are potentially sensitive to short distance modifications of the current.  This matches the our analysis following equation \eqref{modecur}.  Indeed in the long distance limit, the effects of the massive charged particles can be modeled by operator deformations of the Maxwell Lagrangian, and one can adjust the electric current $J_{e}$ to restore conservation.

Let us now consider the short distance limit, $mr \ll 1,$ of the effective potential and the resulting effective charge $q(r)$. As we take the radius of the $S^{2}$ surrounding the Wilson line smaller and smaller, we begin to penetrate the cloud of virtual particles. More generally, we may write \cite{Peskin:1995ev}
\begin{equation}
V(\textbf{x}) = \int \frac{d^3p}{(2\pi)^3} e^{i \textbf{p} \cdot \textbf{x}} \frac{-g^2 q_{\infty}}{|\textbf{p}|^2 [ 1 - \Pi(- |\textbf{p}|^2) ]}~,
\label{eq:potential}
\end{equation}
where $\Pi(p^2)$ is the 1PI modification of the photon propagator due to loops of the charged particle,
\begin{equation}
\Pi_{\mu\nu}(p^2) = \frac{-i\left(\eta^{\mu\nu} - p_\mu p_\nu/p^2\right)}{p^2+i \epsilon} \frac{1}{ 1 - \Pi(p^2)}~,
\end{equation}
with $\Pi(0)=0$.
Expanding the denominator of \eqref{eq:potential} as $1/(1 - \Pi) \simeq 1+ \Pi $, the first term gives the usual $1/r$ Coulomb potential. The term proportional to $\Pi$ gives the correction
\begin{equation}
\delta V(r) =- \frac{g^2q_{\infty}}{2 \pi^2 r}  \int_{2m}^\infty dp \frac{e^{-p r}}{p} \text{Im}[\Pi(p^2) ] =- \frac{g^2}{4 \pi r} \frac{n^2 g^2}{6\pi^2} \int_{2m}^\infty dp \frac{e^{-p r}}{p} \sqrt{1 - \frac{4m^2}{p^2}}\left( 1 + \frac{2m^2}{p^2} \right)~,
\label{eq:deltaV}
\end{equation}
which comes from the branch cut in $\Pi(-|\textbf{p}|^2)$ beginning at $|\textbf{p}| = 2 m i$ due to the charged particle of mass $m$.  We now expand in an series for small $mr$  \eqref{eq:deltaV} and obtain:
\begin{equation}
\delta V(r)/g^{2} \simeq \left( \frac{q_{\infty}}{4 \pi r}\right)\left(\frac{g^{2}n^{2}}{6\pi^{2}}\right) \left(\text{const}+\log(mr)+\mathcal{O}(mr)\right)~.
\label{eq:smallm}
\end{equation}

We can pass from \eqref{eq:smallm} to an expression for the effective charge:
\begin{equation}\label{qvar}
q(r)/q_{\infty}=1-\left(\frac{g^{2}n^{2}}{6\pi^{2}}\right)\log(mr)+g^{2}\left(\text{const}+\mathcal{O}(mr)\right)~.
\end{equation} 
We recover the expected behavior: the electric charge runs logarithmically with distance $r$ for a particle of mass $m \ll 1/r$.  Notice now that in addition to the analytic terms above there is a non-analytic logarithm.  Following our discussion around \eqref{jmod}, this logarithm gives rise to a breaking of the electric one-form symmetry that is scheme independent, i.e. it is independent of how the current operator is modified at short distances.

Let us now investigate the criterion \eqref{eq:GSB} for the distance scale $\rbreak$ where the electric 1-form symmetry is badly broken by screening effects of the charged particles.  As described above the scheme independent symmetry breaking effects come from the logarithms in \eqref{qvar}.  In general, given a spectrum of charged particles of masses $m_{i}$ and integer charges $n_{i}$ the effective variation of $q(r)$ is:
\begin{equation}
\left(\frac{1}{q_{\infty}}\frac{dq(r)}{d\log(r)}\right)\approx \sum_{i|m_{i}r<1}\frac{g^{2}n_{i}^{2}}{6\pi^{2}}~.
\end{equation}
The symmetry breaking scale  $\rbreak$  is where the right-hand side is order one so that the logarithms in the effective charge have an order one coefficient. For a weakly-coupled theory, $g \ll 1$, we see that badly breaking the global symmetry at the scale $r$ requires a light particle ($m \ll 1/r$) with very large charge, $n \sim 1/g$. Alternatively it requires a tower of light charged particles.

We can now extract the typical properties of a particle in the tower by considering averaged quantities.  Define the average charge $\langle n^2 \rangle_{\Lambda}$ of particles up to mass scale $\Lambda$ via:
\begin{equation}
 \sum_{i|m_i < \Lambda} n_i^2 \equiv N(\Lambda) \langle n^2 \rangle_{\Lambda} ~,
\end{equation}
where $N(\Lambda)$ the number of species with mass below $\Lambda$.  Imposing that the electric-one form symmetry is badly violated by the time we arrive at the gravitational cutoff scale $\LQG$ implies:
\begin{equation}\label{eq1}
g^{2} N(\LQG) \langle n^2 \rangle_{\LQG} \gtrsim 1~,
\end{equation}
where we have neglected order one coefficients to focus on the parametric behavior.  The above inequality is essentially equivalent to the weak gravity conjecture.  Indeed, the number of species, $N(\LQG)$, the cutoff $\LQG$, and the Planck mass $M$ are related as:
\begin{equation}\label{eq2}
N(\LQG)\Lambda_{\rm QG}^{2} \sim M_{\textrm{Pl}}^{2}~.
\end{equation}
where the effective cutoff $\LQG$ is lower than the naive Planck mass due to the effects of particles renormalizing the graviton propagator.  We also know that by construction the average value of the mass  must be smaller than the cutoff considered hence $\langle m^2 \rangle_{\Lambda} \lesssim \Lambda^2$.  Combining this with equations  \eqref{eq1} and \eqref{eq2} leads to the weak gravity conjecture bound:
\begin{equation}\label{aWGC}
g^{2}\frac{ \langle n^2 \rangle_{\LQG}}{\langle m^2 \rangle_{\LQG}}\gtrsim \frac{1}{M_{\textrm{Pl}}^{2}}~.
\end{equation}
In other words, the average particle has a superextremal charge-to-mass ratio, up to $O(1)$ factors. This implies that at least one particle is (roughly) superextremal, satisfying the weak gravity conjecture, and it suggests the existence of an infinite tower of superextremal particles, satisfying the tower weak gravity conjecture.\footnote{In a theory with multiple photons, we can run the above argument with respect to any $U(1)$ of our choosing. As a result, we conclude that there must be a tower of (roughly) superextremal particles with respect to any $U(1)$ direction in the charge lattice, which implies that the weak gravity conjecture will be satisfied for each such direction. This in turn implies the Convex Hull Condition \cite{cheung:2014vva}.}

It is instructive to compare our analysis to that of \cite{Heidenreich:2017sim}. There, the same computations were performed, and justified using the idea that the EFT consisting of the photon and charged particles must become strongly coupled by the Planck scale.  This implies that the loop corrections to the photon propagator must rival the tree level result at or below the cutoff scale $\LQG$ leading again to \eqref{aWGC}.  Here we see that the same idea follows directly from a symmetry principle, the breaking of $U(1)_{e}^{(1)}$ by screening effects of light charged particles.  This idea is closely related to the physical picture of the weak gravity conjecture advocated in \cite{Harlow:2015lma}: when the $U(1)_{e}^{(1)}$ is broken by the presence of charged particles, the Wilson lines can end.  The unifying theme in all of these derivations of the weak gravity conjecture is the idea that the fundamental mechanism of, symmetry breaking, or strong coupling, or Hilbert space factorization, must be one that can be modeled in low-energy effective field theory.

\subsubsection{Example: The Kaluza-Klein Photon}

Let us illustrate the above points with a simple example: Kaluza-Klein reduction of pure gravity on a circle.

This example was studied in the context of emergence in \cite{Heidenreich:2017sim}, and our treatment is similar. The reduction of pure gravity in $5$ dimensions on a circle of radius $R$ produces a $U(1)$ gauge field in four dimensions with coupling constant
\begin{equation}
\eKK^2 = \frac{2}{R^2} \frac{1}{\Mpl^{2}} = \frac{1}{\pi R^3} \frac{1}{\MpD^{3}}~,
\end{equation}
where we have used the relationship between the Planck scales across dimensions:
\begin{equation}\label{eq:PlanckdD}
\Mpl^{2}=(2\pi R)\MpD^{3}~.
\end{equation}

For every $n \in \mathbb{Z}$, there is a graviton Kaluza-Klein mode of charge $n$ under the Kaluza-Klein photon whose mass is given by
\begin{equation}
m_n^2 = \frac{n^2}{R^2}\,~,~~~n \in \mathbb{Z}~.
\end{equation}
These charged particles are exactly extremal, saturating the weak gravity conjecture bound \eqref{WGCstate}. Moreover, when these charged particles run in loops they induce corrections to both the graviton propagator (thereby lowering the quantum gravity scale) as well as the gauge propagator (thereby contributing to the breaking of the electric 1-form global symmetry). To be precise, we have
\begin{equation}
\Mpd^{2} = N(\LQG) \LQG^{2} \sim R  \LQG^{3}~,
\end{equation}
using the fact that there are roughly $R \LQG$ graviton Kaluza-Klein modes of mass below $\LQG$. Using \eqref{eq:PlanckdD}, we thus have 
\begin{equation}
\LQG \sim \MpD~.
\end{equation}
In other words, gravity becomes strongly coupled at the Planck scale $\MpD$ of the five-dimensional theory theory rather than the Planck scale of the reduced theory. This is to be expected, since gravity propagates in the full five-dimensional spacetime.

We can similarly compute 1PI loop contributions to the photon propagator from charged particles in order to determine the scale at which the electric 1-form global symmetry is badly broken in the $d$-dimensional theory. We have
\begin{equation}
\Pi(\Lambda^2) \sim \eKK^2 \sum_{n|m_n < \Lambda} n^2  \sim  \frac{\MpD^{3}}{R^3}  (\Lambda R)^3~.
\end{equation}
The 1-form global symmetry is badly broken when this becomes $O(1)$, which happens precisely when $\Lambda \sim \MpD \sim \LQG$: the global symmetry is badly broken right at the quantum gravity scale.

\subsection{Non-Abelian Weak Gravity and Non-Invertible Symmetry}\label{sec:noninv}

In this section we apply the calculations developed so far to the case of weakly-coupled four-dimensional non-abelian gauge theory.  Our aim is to motivate a general non-abelian version of the weak gravity conjecture from symmetry breaking.  Specifically, we will argue that any theory with non-abelian gauge fields coupled to gravity must have sufficiently many matter fields such that at the quantum gravity scale the theory is strongly coupled 
\begin{equation}\label{nabwgc}
	g(\Lambda_\text{QG}) \gtrsim O(1)~.
\end{equation}
We will argue that this condition is needed to badly violate a generalized symmetry defined by non-invertible topological operators.  These are operators that, like more familiar symmetries, do not change under small deformations of their positions, but whose fusion algebra is more general than that of a group.  These non-invertible symmetries have recently been explored in \cite{Bhardwaj:2017xup, Chang:2018iay, Thorngren:2019iar, Rudelius:2020orz, Heidenreich:2021tna, Sharpe:2021srf, Koide:2021zxj, Choi:2021kmx}. Other arguments for conditions similar to \eqref{nabwgc} have appeared in \cite{Heidenreich:2017sim, Grimm:2018ohb, Heidenreich:2018kpg}.

To begin, we first consider pure Yang-Mills theory with a gauge group $K$ at vanishing coupling $g=0$.  We work in conventions where the gauge field $A^{a}$ has canonically normalized kinetic terms, so in the zero coupling limit, the covariant derivative $D = \partial + g A$ becomes the ordinary derivative $\partial$. Thus, the equation of motion is free:
\begin{equation}
\partial_{\mu}{\tilde{F}}^{\mu\nu,a} = 0~,
\end{equation}
where
${\tilde{F}}_{\mu\nu}^{a} = \partial_{\lbrack\mu}A_{\nu\rbrack}^{a}$ is the field strength at zero coupling.

The gauge transformation of the field $A$ is:
\begin{equation}
A(x) \rightarrow k^{- 1}(x)\, A(x)\, k(x) + \frac{1}{g}k^{- 1}(x)\, dk(x)~,
\end{equation}
with $k = e^{i \lambda^{a}(x)T^{a}}$ where $\lambda^{a}(x)$ are the transformation parameters and $T^{a}$ are the generators of the Lie algebra of $K$.  For $A$ to be finite in the $g \rightarrow 0$ limit, the term $k^{- 1}(x)\, dk(x)$ has to be of order $\mathcal{O}(g)$. This condition is achieved when each parameter $\lambda^{a}(x)$ is $\mathcal{O}(g)$. In this case, we set $\lambda^{a} = g{\tilde{\lambda}}^{a}$, and then ${\tilde{\lambda}}^{a}$ behaves as an infinitesimal $U(1)$ (or $\mathbb{R}$) gauge transformation parameter for each component $a$.  Additionally we must also take care of global gauge transformations, i.e., those with constant $\lambda^{a}$ and $k$.  These also preserve finiteness of gauge field $A$ in the zero coupling limit $g \rightarrow 0$.
Therefore, the gauge field $A$ behaves as an $\mathbb{R}^{\dim K}$-gauge field, identified by global $K$ transformations.

As in the case with abelian gauge theory, the equation of motion tells us that ${\tilde{F}}^{\mu\nu,a}$ can be regarded as a current for a continuous one-form symmetry. However, when the gauge group $K$ is non-abelian, the operator ${\tilde{F}}^{\mu\nu,a}$ is not gauge invariant and hence not a valid local operator, even at zero coupling. Therefore, strictly speaking, there is no one-form symmetry generated by ${\tilde{F}}^{\mu\nu,a}$, even at the zero coupling point.

Let us elucidate what is wrong with defining a one-form symmetry by ${\tilde{F}}_{\mu\nu}^{a}$ and making it act on a non-abelian Wilson line. Following equation \eqref{1formabop}, we would define the candidate one-form symmetry generator on a codimension-two surface $\Sigma$ as
\begin{equation}
U_{\alpha,a} = \exp\left( - \frac{\alpha}{g}\int_{\Sigma} \ast {\tilde{F}}^{a}\right)~.
\end{equation}
Meanwhile, the Wilson line $W_{\gamma}(\mathbf{R})$ on the line $\gamma$ in a representation $\mathbf{R}$ of $K$ is defined by the path-ordered exponential: 
\begin{equation}\label{eq:Wnonab}
W_{\gamma}(\mathbf{R}) = P\!\exp\left( i g\int_{\gamma}A^{b}\rho_{\mathbf{R}}(T^{b})\right)~, 
\end{equation}
where $\rho_{\mathbf{R}}$ is the representation map.  To calculate the action of $U_{\alpha,a}$ on $W_{\gamma}(\mathbf{R})$, we choose a gauge with $A^{b}|_{\gamma} = 0$ for $b \neq a$. (This can be achieved by first going to temporal gauge where $A|_{\gamma}$ is constant and then rotating by a global gauge transformation to the $a$ direction. ) With this condition, we can ignore the path-ordering in \eqref{eq:Wnonab},  and compute the action of $U_{\alpha,a}$ on $W_\gamma(R)$ in the $g\to0$ limit as we did in the abelian case. We obtain 
\begin{equation}
  U_{\alpha,a}(S^{d-2}) W_{\gamma}(\mathbf{R}) =  \exp\left( i g\int_{\gamma}A^{a}\rho_{\mathbf{R}}(T^{a}) \right)\times \exp\left( i \alpha \rho_{\mathbf{R}} (T^a) \right)~,
\end{equation}
where the index $a$ on the right hand side is not summed over.  Now we can gauge back the remaining components of $A|_\gamma$, obtaining the general expression without gauge fixing:
\begin{equation}\label{eq:UW_nonab}
  U_{\alpha,a}(S^{d-2}) W_{\gamma}(\mathbf{R}) = 
 P\!\exp\left( ig \int^x A^{b}\rho_{\mathbf{R}}(T^{b})\right) \, e^{i \alpha \rho_{\mathbf{R}} (T^a)(x)} P\!\exp\left( ig \int_x A^{b}\rho_{\mathbf{R}}(T^{b})\right)~.
\end{equation}
Here, $b$ runs over all Lie algebra components.  Thus the effect of the operator $U_{\alpha,a}$ is to insert the matrix $e^{i \alpha \rho_{\mathbf{R}}(T^a)}$ at a point $x\in \gamma$. The point $x$ is arbitrary, but it is natural to identify it as the point where the operator $U_{\alpha,a}$ contracts. (Note that the gauge transformation parameter $\lambda$ we used to reinstate general components of $A$ is of $O(g)$, and thus acts trivially on $T^a$ in the $g\to 0$ limit.). The result \eqref{eq:UW_nonab} is clearly not gauge invariant under global gauge transformations $k(x) = k \in K$, under which $T^a$ is acted on by conjugation.  This means that $U_{\alpha,a}$ itself is not a gauge-invariant operator in the theory.  

We can remedy the problems with $U_{\alpha,a}$ and define a gauge invariant operator by averaging over constant $K$ gauge transformation.  Specifically, consider the following operator:
\begin{equation}\label{U_av}
  U_{\vec\alpha}^\text{av} = \int_K dk \, \exp\left( - \frac{1}{g}\int_{\Sigma} \vec\alpha ( {k^{-1}\ast\tilde{F}}k)\right)~,
\end{equation}
where $\vec\alpha \in \mathfrak{K}^*$ is a linear function on the Lie algebra, (for instance projection onto a given matrix component) and $dk$ is the Haar measure on $K$. This operator is not gauge invariant under a non-constant gauge transformations, but such transformations are irrelevant in the $g\to 0$ limit, since they make $A$ diverge.

We can now combine the calculation \eqref{eq:UW_nonab} together with the averaging definition \eqref{U_av} to determine the action of $U_{\vec\alpha}^\text{av}$ on a Wilson line operator.  We have
\begin{equation}\label{eq:UW_nonab_av}
   U_{\vec\alpha}^\text{av}(S^{d-2}) W_{\gamma}(\mathbf{R}) = 
 P\!\exp\left( ig \int^x A^{b}\rho_{\mathbf{R}}(T^{b})\right) \, \left(\int_K dk\, \rho_{\mathbf{R}} \left(k^{-1}e^{i\vec\alpha^\vee}k\right)\right)  P\!\exp\left( ig \int_x A^{b}\rho_{\mathbf{R}}(T^{b})\right)~.  
\end{equation}
which is manifestly gauge invariant. Here $\vec\alpha^\vee \in \mathfrak{K}$ is the element of the Lie algebra dual to $\vec\alpha$ with respect to the Killing form, and we have used the fact that conjugation commutes with exponentiation.  The expression in the middle of \eqref{eq:UW_nonab_av} above is a matrix acting on $\mathbf{R}$ that is invariant under conjugation by any element $k\in K$.   Assuming $\mathbf{R}$ is irreducible, we then apply Schur's lemma to conclude that it is proportional to the identity 
\begin{equation}\label{schur}
\int_K dk\, \rho_{\mathbf{R}} \left(k^{-1}e^{i\vec\alpha^\vee}k\right)= f_{\mathbf{R}}(\vec{\alpha}) I_{\mathbf{R}}~,
\end{equation}
with $f_{\mathbf{R}}(\vec{\alpha})$ a scalar function and $I_{\mathbf{R}}$ the identity map on $\mathbf{R}$.  As a concrete example, when $K=SU(2)$, $\vec\alpha = \theta (T^3)^\vee$ and $\mathbf{R}$ is the fundamental representation one has 
\begin{equation}\label{UavSU2}
  \int_{SU(2)} dk \, \rho_{\text{fund}}\left(k^{-1} e^{i \theta T^3} k\right) =   \cos(\theta) I~.
\end{equation}
Thus combining \eqref{eq:UW_nonab_av} with \eqref{schur}, we find that the averaged operators $U_{\vec\alpha}^\text{av}$ act on Wilson lines by scalar multiplication: 
\begin{equation}
U_{\vec\alpha}^\text{av}(S^{d-2}) W_{\gamma}(\mathbf{R})=f_{\mathbf{R}}(\vec{\alpha})W_{\gamma}(\mathbf{R})~.
\end{equation}
This result is conceptually very similar to the action of a one-form symmetry on a charged line operator with one key difference: the scalar factor above need not have norm one.  This is the hallmark of a non-invertible symmetry.

From the definition \eqref{U_av}, we see that the argument $\vec\alpha$ entering the definition is only meaningful up to conjugation by elements of $K$.  We can use this conjugation action to deduce that the independent operators of the form of \eqref{U_av} are classified by the Weyl chamber of the Cartan subalgebra of $\mathfrak{K}.$   This Weyl chamber does not have a group structure where the origin acts as the identity and hence we do not expect the operators $U^\text{av}_{\vec{\alpha}}$ form a group. More explicitly, when $K=SU(2)$, the Weyl chamber is the positive real axis labeled by $\theta$ above, and \eqref{UavSU2} indicates
\begin{equation}
  U_{\theta_{1}}^\text{av}U_{\theta_{2}}^\text{av} = \frac12 \left(U_{\theta_{1} + \theta_{2}} + U_{\theta_{1}-\theta_{2}}\right)~.
\end{equation}
For general $K$, the algebra is more involved and will be studied in \cite{toappear}. An algebra of topological operators that involves a sum of terms, as opposed to a unique fusion product, is the hallmark of non-invertible symmetry (see e.g.\ \cite{Bhardwaj:2017xup, Chang:2018iay, Thorngren:2019iar, Sharpe:2021srf, Koide:2021zxj, Choi:2021kmx} for the concept itself and \cite{Rudelius:2020orz, Heidenreich:2021tna} for the applications of the concept in the context of quantum gravity).  Indeed, the same set of operators are considered for the 2+1 and 1+1-dimensional gauge theories in \cite{Nguyen:2021yld, Nguyen:2021naa}.

In summary, the construction above shows that non-abelian Yang-Mills theory at zero gauge coupling has a spectrum of non-invertible topological operators that generalize the electric one form symmetry of abelian Maxwell theory.  To connect these non-invertible topological operators with the weak gravity conjecture, we have to improve the construction of $U^\text{av}_{\vec{\alpha}}$ so that the operator is gauge invariant with small but finite coupling $g$.  A simple-minded procedure is to amend the operator order by order in $g$ so that its correlation functions are gauge invariant up to a given order.  A more top down solution is the following.  Instead of averaging over only constant gauge transformations, we can also average over position-dependent gauge transformations on the surface $\Sigma$:
\begin{equation}\label{U_av_infinite}
  U_{\vec\alpha}^\text{av}(\Sigma) = \int_{k\in \text{Map}(\Sigma,K)} \mathcal{D}k \, \exp\left( - \frac{1}{g}\int_{\Sigma} \vec\alpha ( {k^{-1}\ast\tilde{F}}k)\right)~.
\end{equation}
To tame this infinite-dimensional integral, one requires a cutoff, e.g.\ $||k^{-1}dk|| < \Lambda_\text{cutoff}$  with some appropriate norm.\footnote{To see that \eqref{U_av} and \eqref{U_av_infinite} give the same answer in $g\to0$ limit, note that in the limit only the two-point function of $F$ and $A$ is relevant. Then we can average $F(x)$ over the action of $k(x)$ for each point $x\in\Sigma$, and integrate over $\Sigma$. This gives the same result as \eqref{eq:UW_nonab_av}.} 
This definition does not guarantee that we can safely send $\Lambda_\text{cutoff}\to \infty$ without a divergence, but one can set $\Lambda_\text{cutoff} \sim \Lambda_{QG}$ in the context of effective field theory. Then, the operator \eqref{U_av_infinite} is gauge invariant in the regime of validity of the given effective description.

The operators $U_{\vec\alpha}^\text{av}$ defined by \eqref{U_av_infinite} are not topological for non-zero coupling $g$.  However for small coupling they are close to the exactly topological non-invertible symmetry operators of the zero coupling theory defined by \eqref{U_av}.  Thus, for small coupling $g$, these operators define an approximate non-invertible symmetry.   A natural extension of the no global symmetries hypothesis \cite{Rudelius:2020orz}  suggests that exactly topological non-invertible symmetries are disallowed in quantum gravity.  Extending our general analysis of section \ref{sec:approx} to this context, we expect that at the Planck scale, these generalized approximate symmetries must be badly violated.  This suggests that at the quantum gravity scale, gauge theory must be strongly-coupled, motivating the general claim:
\begin{equation}\label{nabwgc2}
	g(\Lambda_\text{QG}) \gtrsim O(1)~.
\end{equation}
This is an analog of the weak gravity conjecture in the context of non-abelian gauge theory.  Indeed, since pure non-abelian gauge theories are asymptotically free, the only way to achieve the strong-coupling condition \eqref{nabwgc2} is to have sufficiently many charged massive charged particles to change the sign of the beta function and drive the theory to strong coupling.  This agrees with the perspective on non-abelian gauge theory and emergence advocated in \cite{Heidenreich:2017sim}.

It is worth remarking that our argument for the condition \eqref{nabwgc2}, is similar to our general logic of section \ref{sec:TWGC}.  If the effective field theory remains valid up to the quantum gravity scale, it should include effects that violate approximate symmetries. However,  in models where a given effective field theory description breaks down far below the Planck scale, an approximate symmetry can remain valid up to the cutoff where the degrees of freedom change.  An example of this that involves non-abelian gauge theory is the heterotic string.  In this theory the Yang-Mills coupling $g$ and the string coupling $g_{s}$ are related by 
\begin{equation}
g^{2} M_{\mathrm{Pl};10}^{6}\sim g_{s}^{1/2}~.
\end{equation}
Thus, at weak string coupling the low-energy Yang-Mills coupling can be parametrically small in Planck units.  However, the string mass scale $m_{s}$ is also parametrically light in this limit:
\begin{equation}
m_{s} \sim g_{s}^{1/4}M_{\mathrm{Pl};10}~.
\end{equation}
Hence far below the Planck scale, one encounters a tower of charged string states and the effective description as non-abelian gauge theory breaks down.  At the transition the Yang-Mills coupling $g$ is small and the $U_{\vec\alpha}^\text{av}$ operators are still nearly topological.  It is natural to expect that the string states badly break these approximate symmetries at energies below the Planck scale.

\subsection{The Magnetic Weak Gravity Conjecture}\label{ssec:magnetic}

We now turn our attention to the magnetic weak gravity conjecture.  For concreteness we focus on the case of spacetime dimension four.  In this case the magnetic weak gravity conjecture implies that in any weakly-coupled abelian gauge theory that can be consistently coupled to quantum gravity, there must exist a superextremal magnetic monopole.  In equations, letting $n$ be the magnetic charge, and $m_{\text{mon}}$ the monopole mass this reads:
\begin{equation}\label{MWGCeq}
\frac{n/g}{m_{\text{mon}}} \gtrsim \frac{1}{M_{\textrm{Pl}}}~.
\end{equation}
In this section, we motivate this conjecture from the point of view of symmetry breaking.

As discussed around \eqref{emcurrs}, in the absence of charged magnetic monopoles, the $U(1)$ gauge theory has a one-form symmetry $U(1)_{m}^{(1)}$ which measures the magnetic charge of 't Hooft lines.  As in our general discussion in section \ref{sec:approx} we anticipate that this symmetry must be violated below the quantum gravity scale. 

To begin consider again the associated two form current $J_{m}=g\,F/2\pi$. Notice that this operator is closed (the Hodge dual is conserved) by purely topological considerations: i.e. $F=dA$ implies $dF$ vanishes.  In particular this means that as long as the effective field theory contains a $U(1)$ gauge field there will be a conserved magnetic one-form symmetry.  Hence, one cannot violate $U(1)_{m}^{(1)}$ by coupling to charged electric matter, or by adding interactions.  Instead, we must modify the short distance behavior of the theory so that magnetic monopoles become dynamical.

A well-known model with dynamical monopoles is non-abelian gauge theory.  When this theory is Higgsed to an abelian theory, there are massive magnetic monopoles described as solitons.  We here consider the simplest example with non-abelian gauge group $SU(2),$ but analogous comments apply to other examples.   The Lagrangian for this theory is given by
\begin{equation}\label{SU2L}
\mathcal{L} = \frac{1}{2} D_\mu \Phi^a D^\mu \Phi^a - \frac{1}{4} W_{\mu\nu}^a W^{a\,\mu\nu} - \frac{1}{8} \lambda (\Phi^a \Phi^a - \sigma^2)^2~,
\end{equation}
with $G_{\mu}^{a}$ the non-abelian gauge field, $\Phi^a$ a real adjoint-valued scalar field, and
\begin{equation}
W_{\mu\nu}^a = \partial_\mu G_\nu^a - \partial_\nu G_\mu^a + g \epsilon_{abc} G_\mu^b G_\nu^c~,~~~ D_\mu \Phi^a = \partial_\mu \Phi^a + g \epsilon_{abc} G_\mu^b \Phi^c~.
\end{equation}

We assume that the model is weakly-coupled, so the gauge coupling $g$ and the quartic coupling $\lambda$ are small the model can be treated semiclassically.  The potential then gives an expectation value to the scalar:
\begin{equation}
\langle|\Phi|^{2}\rangle = \sigma^{2}~,
\end{equation}
and we assume that this scale is far above any strong coupling scales of the model.  At long distances, this yields a $U(1)$ gauge theory with dynamical 't Hooft-Polyakov monopoles.  

To study the properties of the monopoles in more detail, it is helpful to pick a short distance gauge invariant approximation to the emergent $U(1)_{e}^{(1)}$ and $U(1)_{m}^{(1)}$ currents.  Thus define a two-form operator $J\sim \mathrm{Tr}(\Phi W)$ which flows at long distances to $J_{m}$: 
\begin{equation}\label{Juv}
J \equiv \frac{g}{4\pi \sigma}\mathrm{Tr}\left(\Phi W\right)\rightsquigarrow \frac{g}{2\pi}F=J_{m}~ \Longrightarrow \frac{2\pi i}{g^{2}}*J\rightsquigarrow \frac{i}{g}*F=J_{e}~.
\end{equation}
Of course there are other UV operators that flow to $J_{m}$ and $J_{e}$, but the analysis below is not parametrically sensitive to this choice.

In a monopole configuration, the scalar field $\Phi$ winds around the vacuum manifold at spatial infinity, and we have a solution as $r \rightarrow \infty$ of the form
\begin{equation}
\Phi^a(r,t) \rightarrow \sigma \hat {r}^{a} ~,\hspace{.2in} G_{i}^{a} \rightarrow \frac{1}{gr}\epsilon_{i a b} \hat r_b~,\hspace{.2in}W^{a}_{ij}\rightarrow \frac{1}{gr^{2}}\epsilon_{ijb}\hat{r}_{b}\hat{r}_{a}~.
\end{equation}
where $i = 1, 2, 3$ are the spatial directions. Evaluating the magnetic current $J_{m}$ in this configuration yields
\begin{equation}
J_{m}=\frac{g}{4\pi} F_{ij} = \frac{g}{4 \pi\sigma} \Phi^a W_{ij}^a = \frac{1}{4\pi r^2}  \epsilon_{ijk}\hat r^k~,
\end{equation}
and integrating around the sphere at spatial infinity, we find that this configuration carries unit magnetic charge.  Meanwhile, at finite values of $r$, an appropriate ansatz for the monopole solution is
\begin{equation}
\Phi^a(r,t) = \frac{\hat r}{g r} H(\sigma g r)~,~~ G_i^a(r, t) \rightarrow \epsilon_{i a b} \hat r_b \frac{1 - K(\sigma g r) }{ gr}~,
\end{equation}
and the functions $H$ and $K$ may be found numerically.\footnote{In the special limit $\lambda \rightarrow 0$, there are BPS monopoles which are known exactly: the functions $H(\xi)$ and $K(\xi)$ are then given by
\begin{equation}
H(\xi) = \xi \coth \xi -1~,~~~K(\xi) = \xi \csch \xi~.
\end{equation}
} 
In particular, the characteristic scale of the radial dependence of the monopole solution, is $r_{\text{mon}} \sim 1/(\sigma g)$. Note that this is inverse to the W-boson mass $m_{W}\sim g\sigma$, and thus inside the radius $r_{\text{mon}}$ the non-abelian degrees of freedom are excited.  Since we assume small $g,$ $m_{W}$ is far below the Higgs expectation value $\sigma$.  By contrast, at weak coupling the monopole is a parametrically heavy soliton:
\begin{equation}\label{eqscales}
m_{W}\sim \sigma g\ll \sqrt{ \langle|\Phi|^{2}\rangle }\sim \sigma  \ll m_{mon}\sim \sigma/g~.
\end{equation}
Since the W-boson mass can be interpreted as the effective cutoff scale $\Lambda$ for the validity of the low-energy $U(1)$ gauge theory, the magnetic QGC can alternatively be understood as a constraint on $\Lambda$:
\begin{equation}\label{cuteq1}
\frac{1/g}{m_{\text{mon}}} \gtrsim \frac{1}{M_{\textrm{Pl}}} \longleftrightarrow \Lambda \lsim g M_{\textrm{Pl}}~.
\end{equation}

To motivate \eqref{cuteq1} from the point of view of symmetry breaking, we perform a parallel analysis to that of section \ref{sec:TWGC}.  Specifically, we introduce a line operator which flows at long distances to the magnetically charged 't Hooft line.  In the IR $U(1)$ EFT such a line appears to have an exactly conserved $U(1)_{m}^{(1)}$ charge, and our goal is to measure the violation of this charge due to the non-abelian degrees of freedom of the UV gauge theory.  At the technical level, this means we must measure the charge of the UV operator \eqref{Juv} in the presence of the line.

One natural choice for a UV line operator is the 't Hooft line of the UV $SU(2)$ group.  Instead, we here adopt a different approximation that makes electromagnetic duality more manifest. The Wilson line in an abelian gauge theory can be written as 
\begin{equation}
  W_\gamma(q) = \exp\left(i g q\oint_\gamma A\right) = \exp\left(i g q\int_H dA\right) =  \exp\left(2\pi i  q\int_H J_{m}\right) ~,
\end{equation}
where $H$ is a half-plane bounded by the line $\gamma$ and $J_{m}$ is the magnetic one-form symmetry current in \eqref{emcurrs}.  By taking the electromagnetic dual, we can express the 't Hooft line in $U(1)$ gauge  theory as 
\begin{equation}
  T_\gamma^{\text{IR}}(q) = \exp\left(2\pi i  q\int_H J_{e}\right)~,
\end{equation}
where now $J_{e}$ is the electric one-form symmetry current in \eqref{emcurrs}.  As a microscopic line operator in the $SU(2)$ gauge theory approximating the above we now use an analogous formula but with $J_{e}$ replaced by its short distance completion in \eqref{Juv}.  This gives:
\begin{equation}
T_\gamma^{\text{UV}}(q_{\infty}) \equiv \exp\left(-\frac{4\pi^{2}q_{\infty}}{g^2} \int_H *J\right)=\exp\left(-\frac{\pi q_{\infty}}{g\sigma } \int_H *\mathrm{Tr}\left(\Phi W\right)\right)~.
\end{equation}
Our goal is now to evaluate the scale dependence of the approximate $U(1)_{m}^{(1)}$ charge of this line operator.

The effective charge $q(r)$ is now defined via the correlation function of the approximate current $J$ and the UV 't Hooft operator defined above:
\begin{equation}\label{eq:Tomega}
   \exp\left(i \alpha \int_{S_r^{2}}J\right)T_{\gamma}^{\text{UV}}(q_{\infty}) = \exp(i\alpha q(r)) T_\gamma^{\text{UV}}(q_{\infty}) + \cdots ~,
\end{equation}
which is the analog of the electric calculation of the effective charge performed in \eqref{eq:Wardexp}. As the both symmetry and the charged objects are approximate, the action of the charge operator on $T_\gamma$ is not necessarily exactly diagonal as indicated by $\cdots$ in \eqref{eq:Tomega}.  Furthermore, there is a contact term singularity when two $J$'s collide, which occurs in the integration involved in \eqref{eq:Tomega}. As we explain below, we choose this contact term so that $q(r)\rightarrow q_{\infty}$ as $r\rightarrow \infty$. 

To compute $q(r)$ in more detail, we expand the correlator \eqref{eq:Tomega} in powers of $J$.  This expansion is justified in the weak coupling limit of the UV theory where the only connected diagrams come from two point functions.  In this way can calculate the effective charge from the 2-point function of $J$'s:
\begin{equation}\label{qr_from_jj}
q(r)=\left\langle \left(i\int_{S^2_{r}}J \right)\left( \frac{-4\pi^2 q_{\infty}}{g^2}\int_H *J \right) \right\rangle+O(\lambda, g)~.
\end{equation}

To evaluate the integrals, it is convenient to express the two point function in momentum space.\footnote{We use standard conventions where:
\begin{equation}
\langle \mathcal{A}(p)\mathcal{B}(-p)\rangle=\int d^{4}x~e^{-ip x}\langle \mathcal{A}(x)\mathcal{B}(0)\rangle~.
\end{equation}
} The possible tensor structures in the 2-point function of two-form operators are (see e.g.\ Appendix B of \cite{Cordova:2018cvg}):
\begin{equation}
  \left\langle J_{\mu\nu}(p)J_{\rho\sigma}(-p) \right\rangle = 
  A(p^2)\epsilon_{\mu\nu\rho\sigma} + B(p^2) I_{\mu\nu\rho\sigma}(p) + C(p^2)(\delta_{\mu\rho}\delta_{\nu\sigma} - \delta_{\mu\sigma}\delta_{\nu\rho})~,
  \label{eq:OO}
\end{equation}
where $A(p^2),B(p^2)$ and $C(p^2)$ are functions of $p^2$ and the tensor $I_{\mu\nu\rho\sigma}$ is
\begin{equation}
  I_{\mu\nu\rho\sigma} (p) = p_\mu p_\rho \delta_{\nu\sigma} - p_{\nu}p_{\rho}\delta_{\mu\sigma} - p_{\mu}p_{\sigma}\delta_{\nu\rho} + p_{\nu}p_{\sigma}\delta_{\mu\rho}~.
\end{equation}
Substituting this into \eqref{qr_from_jj}, the effective charge can be expressed as 
\begin{equation}\label{qresult}
	q(r) = \frac{2 q_\infty}{g^2}\int_{-\infty}^{+\infty}dp
	\,(p^2B(p^2)+C(p^2)) \left(r+\frac{i}{p}\right)e^{i p r}~,
\end{equation}
and its derivative with respect to $r$ is
\begin{equation}\label{delq_mag}
	\frac1{q_\infty}\frac{d q(r)}{d \log(r)}
	= \frac{- 2 r^2 }{g^2}\int_{-\infty}^{+\infty}dp
	\,(p^2B(p^2)+C(p^2)) p\, e^{i p r}~.
\end{equation}

If $J$ were the exactly conserved magnetic two-form current in a $U(1)$ gauge theory, i.e.\ \eqref{emcurrs}, then the structure functions are given by:
\begin{equation}\label{sym}
B^\text{exact}(p^2) = \frac{g^2}{2\pi p^2}~, \quad C^\text{exact}(p^2) = 0~.
\end{equation}
Notice that while $B(p^{2})$ is non-analytic, the fact that $C(p^{2})$ is exactly equal to zero, as opposed to another constant, is subject to a contact term ambiguity.  This originates from the fact that our prescription for $q(r)$ from the integrated correlator involves intersecting surfaces and hence is sensitive to contact terms. More, generally for an approximate symmetry current $J$ we will again fix the contact terms so that in the long distance limit, $p\to 0$, $J$ approaches an exactly conserved current including the correct choice of contact terms: 
\begin{equation}\label{BCboundary}
	\lim_{p\to 0} p^2B(p^2) = \frac{g^2}{2\pi}~,\quad \lim_{p\to 0} C(p^2) = 0~.
\end{equation}
With this requirement, the definition \eqref{qr_from_jj} indeed reproduces the desired value of $q_\infty$ for large $r$.\footnote{Let us also comment on the symmetry dual to $J$. Given \eqref{eq:OO}, the two point function of $*J$ takes the same form with the replacements:
\begin{equation}
A(p^{2})\longrightarrow A(p^{2})~, \quad B(p^{2})\longrightarrow -B(p^{2})~, \quad C(p^{2})\longrightarrow p^2B(p^2)+C(p^2)~.
\end{equation}
Thus, using \eqref{qresult}, for $*J$ to correctly produce a long distance constant charge we need $\lim_{p\to 0} C(p^2) = \frac{g^2}{2\pi}$, which is not compatible with the choice of contact term \eqref{BCboundary} made for the magnetic charge.  This means that the electric current should differ from the hodge dual of the magnetic current by a contact term.  This can alternatively be deduced by coupling free Maxwell theory to background fields for its one-form symmetry and using the $U(1)_{e}^{(1)}\times U(1)_{m}^{(1)}$ mixed anomaly.}

Returning now to the $SU(2)$-Higgs theory, the two-point function is (neglecting factors of order unity):
\begin{equation}
  \begin{split}
  \left\langle J_{\mu\nu}(x) J(0)_{\rho\sigma}  \right\rangle
  &= \frac{g^2}{\sigma^2}\left\langle\Tr\Phi W_{\mu\nu}(x) \Tr \Phi W_{\rho \sigma}(0)\right\rangle\\
  &= \frac{g^2}{\sigma^2}\left\langle (\sigma + H(x)) W_{\mu\nu}^3(x) (\sigma + H(0)) W_{\rho\sigma}^3(0)\right\rangle\\
  &= g^2\left\langle W_{\mu\nu}^3(x) W_{\rho\sigma}^3(0)\right\rangle
  + \frac{g^2}{\sigma^2}\left\langle H(x) H(0) \right\rangle \left\langle W_{\mu\nu}^3(x) W_{\rho\sigma}^3(0)\right\rangle + O(g)~,
  \end{split}
\end{equation}
where we expanded have expanded $\Phi$ in the Higgs field $H$ as $\Phi = \sigma + H$ and assumed unitary gauge.
The two-point function of $W^3_{\mu\nu}$ reproduces the behavior of the free theory explained above:
\begin{equation}
	\left\langle W^3_{\mu\nu}(-p) W^3_{\rho\sigma}(p)\right\rangle = 
	-\frac1{p^2}I_{\mu\nu\rho\sigma} + O(\lambda, g)~,
\end{equation}
with the choice of the contact term reproducing $q(r\to\infty) = q_\infty$.  Meanwhile, the contribution from the second term to the position space 2-point function is
\begin{equation}
	 \frac{g^2}{\sigma^2}\left\langle H(x) H(0) \right\rangle \left\langle W_{\mu\nu}^3(x) W_{\rho\sigma}^3(0)\right\rangle=-\frac{g^2}{\sigma^2} \int \frac{d^4k}{(2\pi)^4} \frac{1}{(p-k)^2+m_H^2} \frac{1}{k^2} I_{\mu\nu\rho\sigma}(k)~.
\end{equation}
The non-analytic parts of the relevant structure functions $B(p^{2})$ and $C(p^{2})$ defined in  \eqref{eq:OO} are
\begin{align}
	B|_\text{non-analytic} &= \frac{g^2}{2\pi p^2}
	\left( 1 + \frac{p^2}{4 (2\pi)^2\sigma^2 }\int_0^1 dx \, x^2 \log (x(1-x)p^2 + x m_H^2) \right)~, \\
	C|_\text{non-analytic} &= -\frac{g^2}{4 (2\pi)^3\sigma^2 }\int_0^1 dx \, (x(1-x)p^2 + x m_H^2) \log (x(1-x)p^2 + x m_H^2)~,
\end{align}
where $x$ is a Feynman parameter.  Substituting these into \eqref{delq_mag} gives 
\begin{equation}\label{qdlogres}
	\left | \frac1{q_\infty}\frac{d q(r)}{d \log(r)}\right|
	\sim \frac{1}{r^2 \sigma^2} + O(g,\lambda)~.
\end{equation}

Equation \eqref{qdlogres} is the final result of our calculation.  From this we deduce that the symmetry breaking scale for the approximate magnetic one-form symmetry defined by $J\sim \frac{g}{\sigma}\mathrm{Tr}(\Phi F)$ is, up to order one numbers, the Higgs vev scale $\sigma$.  Demanding that this scale of symmetry violation lies below the Planck scale, we have the condition
\begin{equation}
  \sigma = g \times m_{\text{mon}} \lsim M_{Pl}~,
\end{equation}
which, using \eqref{MWGCeq} and \eqref{eqscales}, is exactly the magnetic weak gravity conjecture \cite{Arkanihamed:2006dz}.

It is worth remarking that our discussion here, in close conceptual analogy with the ordinary weak gravity conjecture, is rooted in the idea that the symmetry violating effects must be captured by effective field theory.  In fact this gives another way to directly motivate the magnetic weak gravity conjecture.  Dynamical magnetic monopoles are needed to satisfy the magnetic version of the completeness hypothesis.  If the lightest of these magnetically charged objects can be described in effective field theory, as opposed to as a black hole, then its effective size must be larger than its implied Schwarzschild radius. This yields \cite{Arkanihamed:2006dz}
\begin{equation}
r_{\text{mon}}\sim\frac{1}{\sigma g} \gsim r_{S} \sim \frac{m_{\text{mon}}}{M_{\text{Pl}}^{2}}~,
\end{equation}
which is again the magnetic weak gravity conjecture.

\section{Shift Symmetry Breaking}\label{sec:SHIFT}

In this section we apply the ideas of section \ref{sec:approx} to low-energy effective field theories with shift symmetries on scalar fields.  We use the absence of approximate shift symmetries to motivate several proposed consistency conditions in theories of quantum gravity.  For simplicity in this section we work in four spacetime dimensions, though much of the analysis holds more generally.

\subsection{The Distance Conjecture }\label{sec:SDC}

Consider a simple theory with a real scalar field $\mu$
\begin{equation}\label{freeshiftL}
\mathcal{L} =  \frac{1}{2} (\partial \mu)^2 -V(\mu)~.
\end{equation}
While the general Lagrangian above admits a potential $V(\mu)$, often one encounters examples with the potential is flat (or nearly so), so that the field $\mu$ is a modulus.  For instance, this may occur in supersymmetric models where the potential is protected from renormalization.  Below we consider this situation of a modulus $\mu$. In this section we focus on the case where $\mu$ is non-compact and our calculations are similar to those in \cite{Heidenreich:2018kpg, Grimm:2018ohb}, while in section \ref{sec:axion}, we investigate the case of compact scalars.  

With vanishing potential, the Lagrangian \eqref{freeshiftL} has a shift symmetry: $\mu\rightarrow \mu+c$.  This symmetry is spontaneously broken by the choice of vacuum state, i.e., an expectation value for the scalar.  With canonical normalization for the field $\mu,$ the shift parameter $c$ has scaling dimension one and it is therefore convenient to introduce a reference scale $f$, which is used to measure field space.

The shift symmetry of the scalar field is encoded by a conserved current $J$ and charged local operators $\mathcal{O}_{q}(x)$:
\begin{equation}
J(x)=if *d \mu(x)~, \hspace{.5in}\mathcal{O}_{q}(x)=\exp\left(\frac{iq \mu(x)}{f}\right)~.
\end{equation}
Here, the charge $q\in \mathbb{R}$ is real-valued since the shift symmetry is non-compact.  The charge $q(r)$ can be read off from the two point function using the Ward identity as before:
\begin{equation}
q(r)\equiv-q_{\infty}\left \langle \int_{S^{3}_{r}}*d\mu(x) \mu(0) \right\rangle~.
\end{equation}
Using the fact that the two point function in the free theory is $\left \langle \mu(x) \mu(0) \right\rangle=(2\pi |x|)^{-2}$, we  find that the charge $q(r)$ is independent of the radius $r$ and equal to the long distance value $q_{\infty}$.  

In a theory of quantum gravity, we expect this shift symmetry to be violated and hence expect our effective field theory to include interactions that give $q(r)$ non-trivial radial dependence.  Of course, one way to violate the shift symmetry is to include a non-vanishing potential.  However, we are specifically interested in the case of a modulus $\mu$ where the zero momentum interactions preserve the shift symmetry.  Thus, we would like to investigate interactions that violate the shift symmetry at non-zero momentum for the $\mu$ field.

Consider a typical class of model where $\mu$ couples to a field whose effective mass is a function of the modulus:
\begin{equation}\label{mupfs}
\mathcal{L} =  \frac{1}{2} (\partial \mu)^2 + i  \bar \psi \slashed D \psi -  m(\mu) \psi  \bar \psi~.
\end{equation}
For concreteness, we have taken the particle to be a fermion.  If $\mu$ acquires an expectation value $\mu_{0}$ with $m(\mu_{0})$ non-zero then at low momentum we can integrate out the fermion $\psi$.  In general the coupling $m(\mu)$ is not translation invariant in $\mu$ and hence this procedure in principle generates a potential that violates the shift symmetry.  Consistent with our assumptions above, we imagine that this induced potential is tuned to vanish, either due to supersymmetry or an explicit adjustment of counterterms.

Consider instead the opposite limit of high energies relative to the mass scale $m(\mu_{0})$.  Fluctuations in the modulus then couple to the fermions through a Yukawa interaction with coupling strength $\left.\frac{dm}{d\mu}\right|_{\mu_{0}}.$  Assuming this coupling is weak, we can then compute the self-energy correction to the modulus field at one loop (below we work up to order one positive numbers): 
\begin{equation}
\Pi(p^{2})=\raisebox{-.45\height}{\includegraphics[width=4cm]{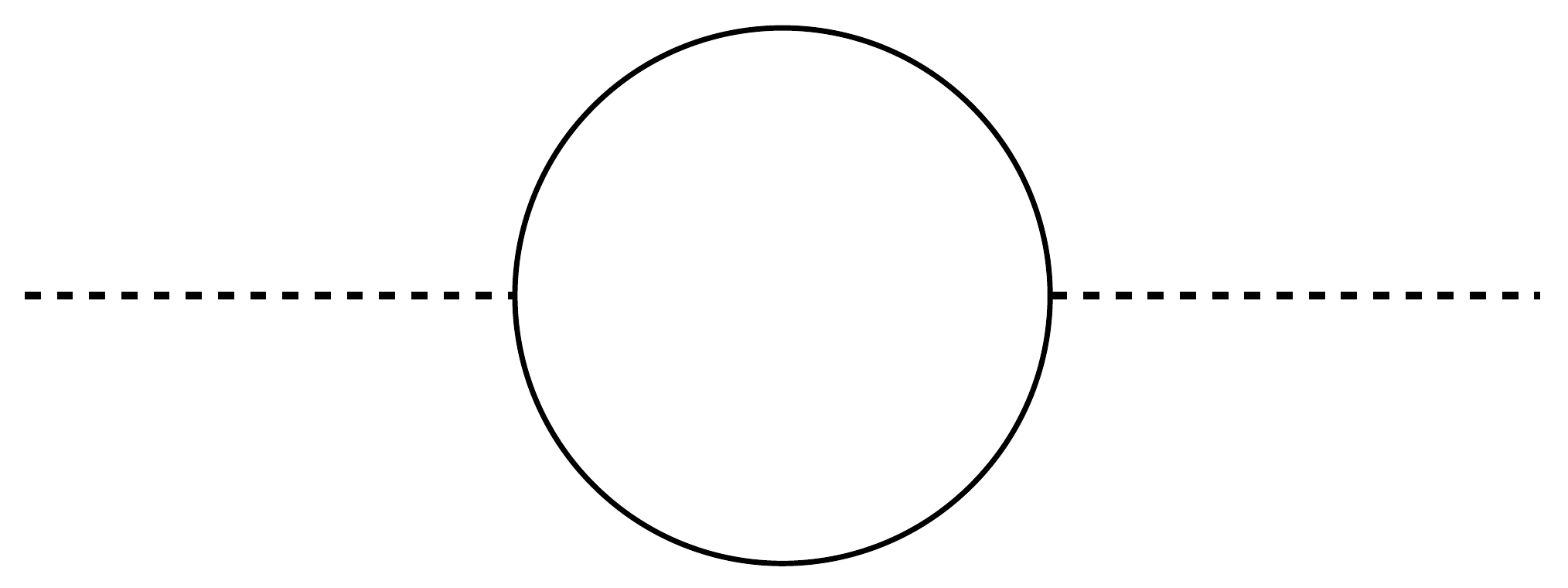}} \approx \left(\left.\frac{dm}{d\mu}\right|_{\mu_{0}}\right)^{2}\log\left(\frac{p^{2}}{m^{2}(\mu_{0})}\right)~, \quad p^{2}\gg m^{2}(\mu_{0})~.
\end{equation}
The two point function of the modulus field is then given by:
\begin{equation}
\left \langle \mu(p) \mu(-p) \right\rangle=\frac{1}{p^{2}(1-\Pi(p^{2}))}\approx \frac{1}{p^{2}}(1+\Pi(p^{2})) \approx \frac{1}{p^{2}}\left(1+ \left(\left.\frac{dm}{d\mu}\right|_{\mu_{0}}\right)^{2}\log\left(\frac{p^{2}}{m^{2}(\mu_{0})}\right)\right)~.
\end{equation}
Equivalently, in position space the short distance approximation to the two-point function is:
\begin{equation}
\left \langle \mu(r) \mu(0) \right\rangle \approx \frac{1}{(2\pi r)^{2}}\left(1-\left(\left.\frac{dm}{d\mu}\right|_{\mu_{0}}\right)^{2}\log\left(m(\mu_{0})r \right)+\cdots \right)~, \quad m(\mu_{0})r\ll 1~,
\end{equation}
where the neglected terms above denote subleading contributions in the small distance limit.  Therefore, at short distances the charged operators have an effective charge
\begin{equation}\label{qsdc}
q(r)/q_{\infty}\approx 1 -\left(\left.\frac{dm}{d\mu}\right|_{\mu_{0}}\right)^{2}\log\left(m(\mu_{0})r \right)~, \quad m(\mu_{0})r\ll 1~.
\end{equation}

Suppose now that we have a spectrum of light fermions coupled to the modulus field $\mu$ with mass functions $m_{i}(\mu)$ as in \eqref{mupfs}.  The logarithmic variation of the effective charge is then a sum over contributions of the form \eqref{qsdc}, where the sum ranges over fermions whose moduli dependent masses are less than a given ultraviolet cutoff $\Lambda$:
\begin{equation}
\left(\frac{1}{q_{\infty}}\right)\frac{dq(r)}{d\log(r)}\approx -\sum_{m_{i}(\mu_{0})< \Lambda}\left(\left.\frac{dm_{i}}{d\mu}\right|_{\mu_{0}}\right)^{2}=- N(\Lambda)\left\langle \left(\left.\frac{dm}{d\mu}\right|_{\mu_{0}}\right)^{2} \right\rangle ~.
\end{equation}
Here, as before, $N(\Lambda)$ denotes the number of light particle species whose masses are lower than $\Lambda$, and the angle brackets denote the average value.  The condition that the shift symmetry of the modulus be strongly violated by the coupling to other fields by the time the cutoff is $\Lambda_{QG}$ is then expressed as an inequality: 
\begin{equation}\label{SCDineq}
 N(\Lambda_{QG})\left\langle \left(\left.\frac{dm}{d\mu}\right|_{\mu_{0}}\right)^{2} \right\rangle \gsim 1~.
\end{equation}

Equation \eqref{SCDineq} is the key result from which one can deduce the properties of the massive $\psi_{i}$ particles as a function of the modulus.  Specifically, let us suppose that as the expectation value of the modulus $\langle \mu \rangle =\mu_{0}$ is sent to infinity, the theory described by the interacting modulus and massive particles becomes parametrically weakly coupled.  In particular, this means that the effective Yukawa couplings must tend to zero as we move to large distance in moduli space.  Therefore we assume:
\begin{equation}
\lim_{\mu \rightarrow \infty}\frac{dm_{i}(\mu)}{d\mu} =0~.
\end{equation}
 If the above weak coupling assumption is valid, it is clear that in order to satisfy the symmetry breaking inequality \eqref{SCDineq}, the number of light species must tend to infinity:
\begin{equation}
\lim_{\mu \rightarrow \infty} N(\Lambda_{QG}) =\infty~.
\end{equation}
In other words, in any parametrically weakly coupled large distance limit in moduli space, a tower of particles must become light to avoid an approximate shift symmetry in the modulus field.  This is a crucial component of the distance conjecture (Conjecture 2 of \cite{Ooguri:2006in}).

In fact we can say more about the behavior of the tower of light particles.  According to \eqref{eq:LQG}, we have:
\begin{equation}\label{mineqmu}
\frac{M^{2}_{\text{Pl}}}{N(\Lambda_{QG})}\sim \Lambda_{QG}^{2} \gsim \langle m^{2}(\mu) \rangle~.
\end{equation}
Since $N(\Lambda_{QG})$ is tending to infinity at large distance in moduli space the effective cutoff $ \Lambda_{QG}$ is also tending to zero for large $\mu$.  This also means that the typical mass in the tower $\langle m^{2}(\mu) \rangle$ is vanishing for large $\mu.$ Moreover, combining \eqref{mineqmu} with \eqref{SCDineq} above we then deduce that:
 \begin{equation}
\left\langle \left(\frac{dm}{d\mu}\right)^{2} \right\rangle \gsim \frac{\langle m^{2}(\mu) \rangle}{M^{2}_{\text{Pl}}}~.
 \end{equation}
Integrating this estimate we deduce that for large modulus $\mu$ we have approximately:
\begin{equation}
\langle|\log(m(\mu))|\rangle  \gsim \frac{\mu}{M_{\text{Pl}}}~.
\end{equation}
So the tower of particles becomes light at least exponentially fast in the moduli distance, with an order one coefficient in Planck units.  This is exactly the more precise version of the distance conjecture of 
\cite{Ooguri:2006in} together with an appropriate estimate on the rate of exponential mass decay \cite{Klaewer:2016kiy}, namely that the coefficient of $\mu$ on the right-hand side above is order one in Planck units.

\subsection{The Axion Weak Gravity Conjecture}\label{sec:axion}

In this section, we consider the axion weak gravity conjecture. Axions are periodic scalar fields, which often arise in string compactifications as the holonomies of $p$-form gauge fields over compact $p$-cycles. They are important for their potential applications to dark matter, inflation, and the strong CP problem \cite{Marsh:2015xka}.

An axion is periodic under $\phi \rightarrow \phi + 2 \pi f$, where $f$ is called the ``axion decay constant.'' In the absence of a potential, the axion transforms under a (spontaneously broken) 0-form shift symmetry as $\phi \rightarrow \phi + c$. Here, we have the periodic identification $c \sim c + 2 \pi f$, which implies that the symmetry group is $U(1)$ rather than $\mathbb{R}$. In the presence of instanton effects, however, $\phi$ typically acquires a periodic potential $V(\phi)$ of the form
\begin{equation}
V(\phi) = \Lambda^4 e^{-S} \cos( \phi /f)~,
\end{equation}
where $\Lambda$ is some UV scale often set by gauge or gravitational dynamics and $S$ is the instanton action. This potential explicitly breaks the 0-form shift symmetry of the axion.

The axion weak gravity conjecture holds that an axion with decay constant $f$ should have an instanton with instanton number $n$ and action $S$ that satisfies
\begin{equation}
\frac{ n }{ f S } \gtrsim \frac{1}{M_{\textrm{Pl}}}~.
\label{eq:0formWGC}
\end{equation}
The magnetic version of the axion weak gravity conjecture holds that an axion string (i.e., a string charged magnetically under $\phi)$ should have a charge-to-tension ratio which is order-one in Planck units or larger. By estimating the tension in terms of the gradient energy of $\phi$ as $T \sim f^2$, \cite{Hebecker:2017wsu, Dolan:2017vmn} argued that the weak gravity conjecture for axion strings takes the form 
\begin{equation}
f \lesssim  M_{\textrm{Pl}} ~.
\label{axionstringWGC}
\end{equation}
In this subsection, we will explain how \eqref{eq:0formWGC} and \eqref{axionstringWGC} can be motivated by demanding that the 0-form shift symmetry of the axion is badly broken below the Planck scale.

As discussed in \cite{Heidenreich:2021yda}, gauge theories with axions can be usefully distinguished by whether or not the gauge field couples to the axion via a $\phi \text{Tr } (F \wedge F)$ coupling. If such a coupling exists, then the electric and magnetic symmetries of the axion and the gauge field are tied up in a higher-group structure \cite{Tanizaki:2019rbk, Hidaka:2020iaz, Hidaka:2020izy, Brennan:2020ehu}, and the excitations of the axion string carry charge under the gauge field \cite{Callan:1984sa}. Kaluza-Klein theory is the prototypical example where such a coupling does not occur.\footnote{In fact, Kaluza-Klein theory may be the only such example, as the work of \cite{Lee:2019wij, Lanza:2021qsu, Heidenreich:2021yda} suggests that the weak coupling limit of a gauge field with a $\theta F \wedge F$ coupling is a tensionless string limit, and a weak coupling limit without such a coupling is a decompactification limit.} With this motivation, we therefore consider the relationship between the axion weak gravity conjecture and the breaking of shift symmetries in two separate cases: (1) axion Yang-Mills theory and (2) axion Kaluza-Klein theory.

\subsubsection{Axion Yang-Mills Theory}

Let us focus first on the case of axion Yang-Mills theory. The Lagrangian is given by 
\begin{equation}
\mathcal{L}  =   \int   \left[\frac{1}{2} d \phi \wedge * d \phi +  \text{Tr} (F \wedge * F) - i P \frac{\phi}{f}  \frac{g^2}{8 \pi^2} \text{Tr} (F \wedge F) \right]~.
\end{equation}
Here, the coupling $P$ is integral to respect periodicity of the axion field.  The instanton action obeys a quantization condition which defines the minimum instanton action $S$ in terms of the gauge coupling:
\begin{equation}
  \int \text{Tr} (F \wedge F) =\frac{8\pi^2 n}{g^2}  ~, \quad n\in \mathbb{Z}\quad\Longrightarrow  S  = \frac{8 \pi^2}{g^2}~.
\label{instantonactioneq}
\end{equation}
As is familiar, instanton effects thus generate a potential for the axion of the qualitative form:
\begin{equation}
V(\phi) = e^{-S} \Lambda^4 \cos(\phi / f) ~.
\end{equation}
This potential breaks the shift symmetry of the axion, but the symmetry-breaking terms are exponentially suppressed for small gauge coupling where $S\gg 1$. They are also power-law suppressed at an energy scale $E$ for $f \gg E$. 

Thus, if the non-abelian gauge coupling remains small at the Planck scale or the axion decay constant $f$ is much larger than the Planck scale, the axion Yang-Mills theory will retain an approximate shift symmetry at the Planck scale.  By contrast if this symmetry is strongly violated by the quantum gravity $E = \LQG$  scale we require:
\begin{align}
f \lesssim  M_{\textrm{Pl}}~,  ~~~~ \text{         and          }~~~~
g(\LQG)\sim 1 \Longrightarrow S(\LQG) \lesssim 1 ~,
\label{implicationeq}
\end{align}
where $g(E)$ and $S(E)$ are respectively, the gauge coupling and instanton action at the energy scale $E$. The first of these conditions is the weak gravity conjecture for axion strings \eqref{axionstringWGC}, while together the conditions of \eqref{implicationeq} are equivalent to the weak gravity conjecture for axions \eqref{eq:0formWGC}.

Notice that the presence of the axion Yang-Mills coupling $\phi \text{Tr} (F \wedge F)$ is crucial for this analysis.  Indeed, this ties the size of the axion potential to the gauge coupling $g$.  According to the general logic of the weak gravity conjecture reviewed in section \ref{sec:TWGC} and the non-abelian analog discussed in section \ref{sec:noninv} gauge couplings must run strong in theories coupled to quantum gravity to break generalized one form symmetries.  This ultraviolet renormalization of the Yang-Mills coupling is driven by a tower of charged particles which drive the gauge sector to strong coupling.  Here we see that this tower of charged particles plays a key role in breaking the axion shift symmetry, similar to the recent analysis of \cite{Heidenreich:2021yda}.

\subsubsection{Axions from Kaluza-Klein Reduction}

A qualitatively different sort of axion model occurs in Kaluza-Klein theory.  In string compactifications, axions arise from integrating $p$-form gauge fields over $p$-cycles of the internal geometry. As a toy model for this phenomenon, we consider the case of an ordinary gauge field in 5d compactified on a circle of radius $R$.  Charged particles in 5d winding around the circle give rise to instantons in 4d, which generate a potential for the axion of the form considered in ``extranatural inflation'' \cite{Arkanihamed:2003wu}.

In this context, the ordinary weak gravity conjecture bound in 5d descends to the both a weak gravity conjecture bound for particles and axion weak gravity conjecture bound for instantons in 4d \cite{Heidenreich:2015nta,Heidenreich:2015wga}. In particular, the axion decay constant $f$ and the instanton action $S$ are given by
\begin{equation}
f=\frac{1}{2 \pi R g}~, \hspace{.5in} S = 2 \pi R m~,
\end{equation}
where above, $m$ is the mass of the 5d particle, and $g$ is the 4d gauge coupling, which is related to the 5d coupling as usual by $g_5^2 = 2 \pi R g^2$.

Now using the fact that $M_{\textrm{Pl}}^2 = 2 \pi R M_{\textrm{Pl}; 5}^3$, we see that the 5d weak gravity conjecture bound implying the existence of a light particle of charge $q$, yields both the 4d weak gravity conjecture bound and the 4d axion weak gravity conjecture bound:
\begin{equation}
 \frac{ g_5  q}{m} \gtrsim \frac{1}{M_{\textrm{Pl}; 5}^{3/2}}~~\Rightarrow~~ \frac{q g}{m} \gtrsim \frac{1}{M_{\textrm{Pl}}}~~\Leftrightarrow~~ \frac{q}{f S} \gtrsim \frac{1}{M_{\textrm{Pl}}}.
 \label{eq:descent}
\end{equation}  
Relatedly, the 5d electric 1-form global symmetry with current $*F^{(5)}$ descends in 4d to both a 1-form global symmetry with current $*F$ and a 0-form global symmetry with current $*d \phi$. At energies larger than the KK scale $1/R$ but smaller than the 5d quantum gravity scale $\LQG$, we have a well-behaved effective field theory description in five dimensions. Demanding that the 5d 1-form global symmetry is badly broken below the 5d Planck scale, the argument of section \ref{sec:TWGC} implies the existence of a tower of superextremal particles in 5d, which descend via \eqref{eq:descent} to a superextremal instantons in 4d, thus implying the axion weak gravity conjecture.

\section{Conclusions}\label{sec:CONC}

We have seen that a number of quantum gravity conjectures including the weak gravity conjecture, the distance conjecture, the magnetic and axion versions of the weak gravity conjecture, can be motivated, and in a sense unified, under the assumption that global symmetries should be badly broken within the context of effective field theory. In the case of the weak gravity conjecture, we were in fact able to motivate the existence of a tower of states of superextremal particles, which is closely-related to the tower and sublattice forms of the weak gravity conjecture.

Our analysis has left us with a number of other open questions and interesting directions for future research. First of all, is there a stronger argument to be made for the statement that global symmetries should be badly broken within effective field theory, below the scale at which quantum gravity becomes strongly coupled? Previous arguments against global symmetries in quantum gravity have been given in the context of black hole physics and AdS/CFT: can these arguments be made more precise, so as to quantify the degree of global symmetry breaking required in a consistent theory of quantum gravity?

The arguments we have given above do not lead us to precise inequalities with $O(1)$ factors included, but rather to approximate versions of the weak gravity conjecture with $\geq$ signs replaced by $\gtrsim$ signs. A more precise definition of ``approximate'' symmetry and ``bady-broken'' symmetry could perhaps lead to more precise inequalities. This would be especially beneficial in the case of the axion weak gravity conjecture and the distance conjecture, whose precise $O(1)$ factors remain a matter of debate.

We have focused exclusively on the breaking of approximate \emph{continuous} global symmetries: it is natural to wonder about approximate \emph{discrete} global symmetries as well. Likewise, we have considered quantum field theories in flat space: it would be interesting to see if any subtleties arise for quantum gravity theories in more general spacetimes, such as dS or AdS.

There are also a number of other versions of the weak gravity conjecture that we have not addressed in this work. For instance, we have not examined the weak gravity conjecture for $p$-form gauge fields with $p>1$, which lead to $p$-form electric symmetries and $(d-p-2)$-form magnetic symmetries that must be broken by charged $(p-1)$-branes and $(d-p-3)$-branes, respectively. Relatedly, we have not considered the topic of higher-group symmetries \cite{Cordova:2018cvg, Benini:2018reh, Tanizaki:2019rbk, Brennan:2020ehu, Cordova:2020tij, Brauner:2020rtz, DelZotto:2020sop, Hsin:2021qiy}, which lead to mixing of weak gravity conjectures between $p$-form gauge fields of different degrees  \cite{Heidenreich:2021yda, Kaya:2022edp}. Similarly, we have not delved into the breaking of so-called ``$(-1)$-form symmetries'' \cite{Cordova:2019jnf, Cordova:2019uob, Tanizaki:2019rbk, Heidenreich:2020pkc}, which are related to the absence of free parameters in quantum gravity and the absence of nontrivial cobordism groups \cite{McNamara:2019rup, McNamara:2020uza}. It would be interesting to connect these conjectures more directly to the breaking of approximate global symmetries, or perhaps even to connect global symmetry breaking to additional generic features of quantum gravity theories that have not yet been discussed in the literature.

The last century of theoretical physics has repeatedly demonstrated the power of symmetry. In this work, we have seen that this power extends to the weak gravity conjecture, which gives us hope that symmetry (and the absence of symmetry) will play an increasing role in our understanding of quantum gravity.

\section*{Acknowledgements}

We thank Daniel Harlow, Ben Heidenreich, Emil Martinec, Hirosi Ooguri, and Matthew Reece for useful discussions.
CC is supported by the US Department of Energy DE-SC0021432 and the Simons Collaboration on Global Categorical Symmetries.
KO is supported by the Simons Collaboration on Global Categorical Symmetries.
TR is supported by the Berkeley Center for Theoretical Physics; by the Department of Energy, Office of Science, Office of High Energy Physics under QuantISED Award DE-SC0019380 and under contract DE-AC02-05CH11231; and by the National Science Foundation under Award Number 2112880.
We acknowledge hospitality from several institutions where portions of this work were completed: the Institute for Advanced Study in Princeton, NJ; the Amherst Center for Fundamental Interactions at UMass Amherst, site of the workshop ``Theoretical Tests of the Swampland.''

\bibliography{ref}
\bibliographystyle{utphys}
\end{document}